\documentclass[11pt,a4paper]{article}
\usepackage{app/jheppub}
\usepackage{bm}
\usepackage{makecell}
\usepackage{float}

\usepackage{epstopdf}

\newcommand{\bB}{{\bm{B}}}

\newcommand{\mR}{{\mathcal{R}}}

\title{\boldmath
Weak Decays of Antitriplet Charmed Baryons from the Perspective of Flavor Symmetry}

\author{Huiling Zhong,} 
\author{Fanrong Xu,}
\author{Qiaoyi Wen, Yu Gu}

\affiliation
{Department of Physics, Jinan University,\\
$\;$Guangzhou 510632, P.R. China}

\emailAdd{fanrongxu@jnu.edu.cn}

\abstract{
In this work, we carry out a global fit for the two-body 
weak decays of antitriplet charmed baryons in both
SU(3) respected and broken scenarios incorporating all the available data up to date.
In the SU(3) irreducible representation approach (IRA), more amplitudes for irreducible
representation terms are taken into account and the ranges for their coefficients 
in each scenarios are predicted. 
By a comparison among various fitting schemes in this work, experimental data prefer the
SU(3) symmetry breaking scenario.
Observables of interest, branching fractions and
decay asymmetries of all the Cabibbo-favored (CF), singly Cabibbo-suppressed (SCS)
and doubly Cabibbo-suppressed (DCS) channels are calculated in the chosen fitting
scheme.  Most of our predictions are consistent well
with experimental data. 
We further propose more ways to explore
the SU(3) symmetry in charmed baryon decays:
(i) A clear measurement of branching fraction on CF mode $\Xi_c^0\to \Xi^0\pi^0$ 
as a large difference exists between SU(3) respected and broken scenarios. 
(ii) The decay asymmetries of the five yet-to-be measured CF  
decay modes, 
$\Xi_c^0\to \Lambda^0 K_S, \Sigma^0 K_S, \Xi^0 \pi^0, \Xi^0 \eta, \Xi^0 \eta'$, for their 
opposite signs in the two different cases.
(iii) The future measurement on  branching fraction ratios indicated from
Eq. (\ref{eq:SU3}) and presented on Table \ref{tab:prediction2}.
An improved measurement for $\alpha(\Lambda_c^+\to p K_S)$ 
is  called for since predictions from
both fittings, including the one in current work and earlier works, and the pole model calculation 
prefer an opposite sign from previous experimental value.
Given branching fractions of most of the CF modes and part of the SCS modes 
being measured,  predictions for the remaining channels, including the DCS modes
as well as more decay asymmetries, are anticipated to be checked by the upcoming 
experiments in BESIII, Belle/Belle-II and LHCb.
}


\begin{document}
\maketitle
\flushbottom


\newpage

\section{Introduction}
\label{sec:Intro}

In recent years,  more and more attentions have been paid on
two-body hadronic weak decays of charmed baryons 
both experimentally and theoretically.  Especially for the branching fractions
and decay asymmetries of antitriplet charmed baryon decays,  
continuous progresses have been made in summer 2022. 
For the first time, branching fractions of the following channels have been measured by BESIII \cite{BESIII:2022bkj, BESIII:2022izy,BESIII:2022tnm,BESIII:2022wxj}
\begin{equation}
\begin{split}
&\mathcal{B}(\Lambda_c^+\to n\pi^+)=(6.6\pm 1.2\pm0.4)\times 10^{-4},\\
&\mathcal{B}(\Lambda_c^+\to p \eta')= (5.62^{+2.46}_{-2.04}\pm 0.26)\times 10^{-4}, \\
&\mathcal{B}(\Lambda_c^+\to \Lambda^0 K^+)= (6.21\pm 0.44\pm0.26\pm0.34)\times 10^{-4} ,\\
&\mathcal{B}(\Lambda_c^+\to\Sigma^+ K_S^0)= (4.8\pm 1.4\pm0.2\pm0.3)\times 10^{-4},\\
&\mathcal{B}(\Lambda_c^+\to\Sigma^0 K^+)= (4.7\pm 0.9\pm0.1\pm0.3)\times 10^{-4}.
\end{split}
\end{equation}
Meanwhile, Belle has made independent contributions in the measurements of both branching fractions and 
decay asymmetries, giving
\cite{Belle:2021vyq, Belle:2022uod, Belle:2022bsi}
\begin{equation}
\begin{split}
&\mathcal{B}(\Lambda_c^+\to p \eta')= (4.73\pm 0.82\pm0.47\pm0.24)\times 10^{-4}, \\
&\mathcal{B}(\Lambda_c^+\to \Lambda^0 K^+)= (6.57\pm 0.17\pm0.11\pm0.35)\times 10^{-4} ,\\
&\mathcal{B}(\Lambda_c^+\to\Sigma^0 K^+)= (3.58\pm 0.19\pm0.06\pm0.19)\times 10^{-4},\\
&\mathcal{B}(\Lambda_c^+\to \Sigma^+ \eta)= (3.14\pm 0.35\pm0.11\pm0.25)\times 10^{-3}, \\
&\mathcal{B}(\Lambda_c^+\to \Sigma^+ \eta')= (4.16\pm 0.75\pm0.21\pm0.33)\times 10^{-3} ,
\end{split}
\end{equation}
and 
\begin{equation}
\begin{split}
&\alpha(\Lambda_c^+\to \Lambda^0 K^+ )= -0.585\pm 0.049\pm 0.018, \\
&\alpha(\Lambda_c^+\to \Sigma^0 K^+)= -0.55\pm 0.18 \pm 0.09,\\ 
&\alpha(\Lambda_c^+\to \Sigma^+ \eta)= -0.99\pm0.03\pm0.05,\\ 
&\alpha(\Lambda_c^+\to \Sigma^+ \eta')=-0.46\pm 0.06\pm 0.03.
\end{split}
\end{equation}
In addition,  the branching fraction measurements of $\Lambda_c \to \Sigma^+ K_S$
and $\Lambda_c\to p \eta'$ have been improved by BESIII and Belle, respectively. 
Moreover, decay asymmetries of $\Lambda_c^+\to \Lambda^0 \pi^+, \Sigma^0 \pi^+$ and
$\Sigma^+\pi^0$ have been measured more precisely.

The dynamics of  charmed baryon weak decays, especially for antitriplet charmed baryon, 
had a long research history for several 
decades \cite{Korner:1992wi, Xu:1992sw, Cheng:1993gf, Ivanov:1997ra}
and recently  has been studied extensively \cite{Cheng:2018hwl, Zou:2019kzq, Niu:2020gjw, Groote:2021pxt} 
promoted by rapid developments in experiment, which has
been reviewed in \cite{Cheng:2021qpd}.
These  phenomenological studies are model dependent, 
due to the lack of a consistent 
QCD-inspired framework to deal with charmed baryon decays so far,  
still require further checks from experiments.
On the other hand, without involving the non-perturbative details, the fitting methodology provides a tool as
a complementary way to interpret data and further predict the yet-to-be measured channels.  
Two different ways, IRA (irreducible representation approach) and 
TDA (topological diagram approach), have been developed to carry out the fitting analysis. 
In IRA, different decay channels can be connected \cite{Lu:2016ogy, Geng:2019xbo, Huang:2021aqu} via
irreducible representations for the SU(3) flavor symmetry.
Another attempt, TDA,  is to take  topological diagrams as basic elements to implement the 
fitting scheme \cite{Zhao:2018mov}.   
The topic that whether an equivalence between the two approaches exists or not, 
has  been discussed recently for mesons \cite{Wang:2020gmn,  He:2018php, He:2018joe} and for baryons \cite{Hsiao:2021nsc}, and further efforts are still required. 

In this work, we adopt the IRA to carry out such a global fitting analysis. 
Meanwhile in
charmed baryon decays whether the SU(3) flavor symmetry is strictly respected or broken 
is still a question worth answering.
Hence the two scenarios, SU(3) symmetry being kept strictly and broken,
are both taken in the following study  by incorporating the latest  experimental data.  
We will give explicitly the relations among different channels at the amplitude level.
Coefficients of amplitudes in terms of irreducible representations will be fitted after comparing with different fitting schemes.  Branching fractions and decay asymmetries of all the two-body decays, including 
CF, SCS and DCS channels, will be predicted
based on the coefficients in the fixed scheme. 
Armed with the explicit predicted values, 
we propose ways to discriminate SU(3) symmetry between its keeping and breaking scenarios, including the focus on branching fraction of special 
modes ($\Xi_c^0 \to\Xi^0 \pi^0$ as an example), decay asymmetries of particular modes
(like $\Xi_c^0\to \Lambda^0 K_S $) as well as branching fraction ratios of some
certain channels ( of which an example is $\mathcal{B}(\Xi_c^0 \to p\pi^-)/\mathcal{B}(\Xi_c^0 \to p K^-)$).
We point out that theoretical predictions
from both fittings and model calculations as well as experimental measuring  provide
complementary information  to explore charmed baryon weak decays from different aspects. 
In general,  opportunities are indicated in these inconsistent channels, of which an example
is $\alpha(\Lambda_c^+\to p K_S)$.
By combing predictions
in different independent ways, some suggestions for experiments are also given.

The remaining part of this paper is organized as follows. 
In Section \ref{sec:setup}, we set up the framework of the entire fitting work, including theoretical formalism and a summary of experimental data. In both scenarios, connections among decay amplitudes are also presented in this section.  Then fitting schemes are introduced in the first part of Sec. \ref{sec:ana}, 
while in the second part we present all the numerical results and related discussion. 
A conclusion is made in Sec. \ref{sec:con}. For more details of choosing fitting schemes, 
results for a comparison are displayed in Appendix \ref{app:scheme}.

\section{Formalism}
\label{sec:setup}

\subsection{The irreducible representation and kinematics}
The state-of-the-art effective Hamiltonian  approach is adopted to describe charmed baryon decays.
At quark level, the effective Hamiltonian to depict  quark decay process  $c\to \bar{q}_1 q_2 u$ is
\begin{equation}
\begin{split}
&\mathcal{H}_{\rm{eff}}=\frac{G_F}{\sqrt{2}}V_{q_1 c}^* V_{uq_2}(c_+\mathcal{O}_+ +c_-\mathcal{O}_-)+h.c.\\
& \mathcal{O}_+ =\frac12\left(\mathcal{O}_1+\mathcal{O}_2\right),\quad   
\mathcal{O}_- =\frac12\left(\mathcal{O}_1-\mathcal{O}_2\right)
\end{split}
\end{equation}
in which 
$\mathcal{O}_1 = (\bar{q}_1 c)(\bar{u} q_2),  \mathcal{O}_2 = (\bar{u} c)(\bar{q}_1 q_2)$, and the 
bilinear operator is defined as
$(\bar{q}_1 q_2) =\bar{q}_1\gamma^\mu(1-\gamma_5) q_2 $. 
In particular, the quark flavors of the four-quark operator  are denoted as  $(q_1, q_2)= (s, d), \{ (s, s), (d,d)\}, (d, s)$, 
corresponding to Cabibbo-favored (CF),
singly Cabibbo-suppressed (SCS) and doubly Cabibbo-suppressed (DCS) processes, respectively.

From the point of view of SU(3) flavor symmetry,
the four-quark operators $\mathcal{O}_{\pm}$ in the weak Hamiltonian 
fall into two irreducible representations $(\bm{6},\overline{\bm{15}})$
\cite{Savage:1989qr, Geng:2019xbo, Huang:2021aqu},
given by
\begin{equation}
\begin{split}
& H(\overline{15})_k^{ij}=\left(
\left(\begin{array}{ccc}
0& 0 & 0\\ 0 & 0& 0\\ 0 & 0& 0
\end{array}\right),\quad
\left(\begin{array}{ccc}
0& s_c & 1\\ s_c & 0& 0\\ 1 & 0& 0
\end{array}\right),\quad
\left(\begin{array}{ccc}
0& -s_c^2 & -s_c\\ -s^2_c & 0& 0\\ -s_c & 0& 0
\end{array}
\right)\right),\\
& H(6)_{ij}=\left(\begin{array}{ccc}
0 & 0 & 0 \\
0 & 2 & -2s_c\\
0 & -2s_c & 2s_c^2
\end{array}\right)
\end{split}
\end{equation}
with $s_c=\sin\theta_c=\sqrt{0.23}$ \cite{Workman:2022ynf} and 
$(i, j, k) = 1, 2, 3$, in which traceless $H(\overline{{15}})$ is  symmetric 
for the superscript indices while  $H({{6}})$ is symmetric for the 
subscript indices. 
For the two-body decays of antitriplet charmed baryon $B_c$ into hadrons  in the baryon and 
meson octets $B_n$ and $M$, 
the amplitudes are separated as $S$- and $P$-waves \cite{Geng:2019xbo}, given
\begin{equation}
\mathcal{M}=\langle M \bm{B}_n |\mathcal{H}_{\rm{eff}}|\bm{B}_c\rangle
= i\bar{u}_f(A-B\gamma_5) u_i.
\end{equation}
If the $S$- and $P$-wave amplitudes ($A$ and $B$ terms) are  invariant under
SU(3) symmetry, generically  they
can further be parameterized by the irreducible 
representations of SU(3) group,
\begin{eqnarray}
A_0 
&=& a_0 H(6)_{ij}(\bB'_c)^{ik}(\bB_n)^j_k (M)^\ell_{\ell}
+a_1H(6)_{ij}(\bB'_c)^{ik}(\bB_n)^\ell_k(M)^j_\ell
+ a_2 H(6)_{ij} (\bB'_c)^{ik}(M)^\ell_k(\bB_n)^j_\ell \nonumber\\
&&+ a_3 H(6)_{ij}(\bB_n)^i_k (M)^j_\ell (\bB'_c)^{k\ell} +
a'_0  (\bB_n)^i_j (M)^\ell_\ell H(\overline{15})_i^{jk}(\bB_c)_k
+a_4 H(\overline{15})_k^{\ell i}(\bB_c)_j (M)_i^j(\bB_n)_\ell^k \nonumber\\
&& + a_5 (\bB_n)^i_j (M)^\ell_i H(\overline{15})_\ell^{jk}(\bB_c)_k
+a_6 (\bB_n)^j_i (M)_\ell^m H(\overline{15})_m^{\ell i}(\bB_c)_j
+a_7 (\bB_n)_i^\ell (M)^i_j H(\overline{15})_\ell^{jk}(\bB_c)_k\nonumber\\
B_0 &=& A_0\Big|_{a_i\to b_i},
\label{eq:ampSU3}
\end{eqnarray}
where
\begin{equation}
\begin{split}
&\bm{B}_c=\left(\Xi_c^0, -\Xi_c^+, \Lambda_c^+\right),\\
&\bm{B}_n=\left(\begin{array}{ccc}
\frac{1}{\sqrt{6}}\Lambda^0 + \frac{1}{\sqrt{2}}\Sigma^0 & \Sigma^+ & p \\
\Sigma^- & \frac{1}{\sqrt{6}}\Lambda^0 - \frac{1}{\sqrt{2}}\Sigma^0 & n \\
\Xi^-& \Xi^0& -\sqrt{\frac23}\Lambda^0 \end{array}\right), \\
& M=\left(\begin{array}{ccc}
\frac{1}{\sqrt{2}}(\pi^0+ c_\phi \eta + s_\phi \eta') & \pi^+ & K^+ \\
\pi^- & \frac{1}{\sqrt{2}}(-\pi^0+ c_\phi \eta + s_\phi \eta') & K^0 \\
K^- & \overline{K}^0 & -s_\phi \eta + c_\phi \eta'
\end{array}\right),
\end{split}
\end{equation}
and $(\bB'_c)^{ij}\equiv \epsilon^{ijk}(\bB_c)_k$. 
Here the mixing of $\eta$ and $\eta'$ have also been considered in which
$(c_\phi, s_\phi)=(\cos\phi, \sin\phi)$ with $\phi=39.3^\circ$ \cite{Feldmann:1998vh,Feldmann:1998sh}.
The coefficients $a_i$ and $b_i$ will be determined by experimental data.
Thanks to the more and more accumulated data, here we  incorporate the
terms characterized by $a_0', a_{4,5,7}$ (and the
corresponding terms in $P$-wave), which were
neglected in \cite{Geng:2019xbo} and recently have been
recollected in \cite{Huang:2021aqu}.

It is known that in D meson sector, SU(3) breaking effect takes up a certain contribution
in the decays \cite{Cheng:2010ry, Li:2012cfa, Cheng:2021yrn}. 
To incorporate the analogous SU(3) 
breaking effect in the decays of charmed baryons is
a reasonable and interesting attempt. In the earlier studies, it has been shown that
there are several SU(3) representations to describe the breaking effect \cite{Savage:1991wu, Pirtskhalava:2011va}. In this work, we continue to take the assumption that 
$\overline{3}$ is the dominant source of SU(3) breaking \cite{Geng:2018bow}
and the corresponding amplitudes can be parameterized  as
\begin{eqnarray}
A' &=&  u_1 (\bB_c)_i H(\overline{3})^i (\bB_n)^j_k (M)^k_j 
+ u_2 (\bB_c)_i H(\overline{3})^j (\bB_n)^i_k (M)^k_j  \nonumber\\
&&
+ u_3 (\bB_c)_i H(\overline{3})^j (\bB_n)^k_j (M)^i_k 
\nonumber \\
B' &=& A'\Big|_{ u_i\to v_i}
\label{eq:ampSU3b}
\end{eqnarray}
with 
\begin{equation}
H(\overline{3})=(s_c, 0, 0),
\end{equation}
which manifests in the SCS processes.\footnote{
Although this occurs under the $\overline{3}$ dominance assumption, 
the feature is in agreement with
the fit in terms of topological diagrams  \cite{Cheng:2010ry, Cheng:2021yrn}.
} 

Then 
in terms of
the total $S$- and $P$-wave amplitudes 
$ A=A_0+A', B=B_0 + B'$, 
the observables of interest, decay width $\Gamma$ and decay asymmetry $\alpha$,
are given as
\begin{equation}
\begin{split}
&\Gamma = \frac{p_c}{8\pi}\left(\frac{(m_i+m_f)^2-m_P^2}{m_i^2}|A|^2
+ \frac{(m_i-m_f)^2-m_P^2}{m_i^2}|B|^2\right)\\
& \alpha=\frac{2\kappa {\textrm{Re}}(A^* B)}{|A|^2+\kappa^2 |B|^2}
\end{split}
\label{eq:kin}
\end{equation}
where $m_i, m_f, m_P$ stand for the masses of initial state baryons, final state
baryons and mesons with $p_c$ being the c.m. three-momentum in the rest
frame of initial baryon.  And the auxiliary parameter $\kappa$ is 
defined as $\kappa=p_c/(E_f+m_f)=\sqrt{(E_f-m_f)/(E_f+m_f)}$.

\subsection{Decay amplitudes}

\begin{table}[t]
\footnotesize{
\caption{The S-wave amplitudes for Cabibbo-favored decays of $\bB_c\to\bB_n M$ in
terms of coefficients of irreducible representation amplitudes (IRA). 
The P-wave amplitudes can  be obtained by replacing $a_i$ by $b_i$ accordingly.} 
\label{tab:amp-cf}
\vspace{-0.4cm}
\begin{center}
\renewcommand\arraystretch{1.8} 
\resizebox{\textwidth}{!} 
{
\begin{tabular}
{| l c  | l c |}
\hline
Channel & $A$ & Channel & $A$ \\
\hline

$\Lambda_c^+\to \Lambda^0 \pi^+$ & $\frac{\sqrt{6}}{6}(-2a_1-2a_2-2a_3+a_5-2a_6+a_7)$ &
$\Xi_c^+\to \Xi^0 \pi^+$ & $-2a_3-a_4-a_6$\\
\hline

$\Lambda_c^+\to p \overline{K}^0$ & $-2a_1+a_5+a_6$ &
$\Xi_c^0\to \Lambda^0 \overline{K}^0$ & $\frac{\sqrt{6}}{6}(-4a_1+2a_2+2a_3-2a_5+a_6+a_7)$ \\
\hline

$\Lambda_c^+\to \Sigma^0 \pi^+$ & $\frac{\sqrt{2}}{2}(-2a_1+2a_2+2a_3+a_5-a_7)$ &
$\Xi_c^0\to \Sigma^0 \overline{K}^0$ & $\frac{\sqrt{2}}{2}(-2a_2-2a_3+a_6-a_7)$ \\
\hline

$\Lambda_c^+\to \Sigma^+ \pi^0$ & $\frac{\sqrt{2}}{2}(2a_1-2a_2-2a_3 - a_5 + a_7)$ &
$\Xi_c^0\to \Sigma^+ K^-$ & $2a_2+a_4+a_7$ \\
\hline

$\Lambda_c^+\to \Sigma^+ \eta$  & \makecell[c]{$\frac{\sqrt{2}}{6}c_{\phi}(-12a_0-6a_1-6a_2+6a_3+6a'_0 + 3a_5 + 3a_7)$\\$+s_{\phi}(2a_0-a'_0-a_4)$} &
$\Xi_c^0\to \Xi^0 \pi^0$ & $\frac{\sqrt{2}}{2}(-2a_1+2a_3+a_4-a_5)$ \\
\hline

$\Lambda_c^+\to \Sigma^+ \eta'$  & \makecell[c]{$\frac{\sqrt{2}}{6}s_{\phi}(-12a_0-6a_1-6a_2+6a_3+6a'_0+3a_5+3a_7)$\\$-c_{\phi}(2a_0-a'_0-a_4)$} &
$\Xi_c^0\to \Xi^0 \eta$ & \makecell[c]{$\frac{\sqrt{2}}{6}c_{\phi}(12a_0+6a_1-6a_3+6a'_0+3a_4+3a_5)$\\$+\frac{1}{3}s_{\phi}(-6a_0-6a_2-3a'_0-3a_7)$} \\
\hline

$\Lambda_c^+\to \Xi^0 K^+$  & $-2a_2+a_4+a_7$ &
$\Xi_c^0\to \Xi^0 \eta'$ & \makecell[c]{$\frac{\sqrt{2}}{6}s_{\phi}(12a_0+6a_1-6a_3+6a'_0+3a_4+3a_5)$\\$-\frac{1}{3}c_{\phi}(-6a_0-6a_2-3a'_0-3a_7)$} \\
\hline

$\Xi_c^+\to \Sigma^+ \overline{K}^0$ & $2a_3-a_4-a_6$ &
$\Xi_c^0\to \Xi^- \pi^+$ & $2a_1+a_5+a_6$ \\
\hline
\end{tabular}
}
\end{center}
}
\end{table}


\begin{table}[h!]
\caption{The same as Table \ref{tab:amp-cf}, except for singly Cabibbo-suppressed (SCS) decays of $\bB_c\to\bB_n M$.} 
\label{tab:amp-scs}
\vspace{-0.4cm}
\begin{center}
\renewcommand\arraystretch{1.5} 
\resizebox{\textwidth}{!} 
{
\begin{tabular}
{| lc | lc |}
\hline
Channel & $s_c^{-1} A$ & Channel & $s_c^{-1} A$ \\
\hline

$\Lambda_c^+\to \Lambda^0 K^+$ & \makecell[c]{$\frac{\sqrt{6}}{6}(2a_1-4a_2+2a_3+3a_4-a_5+2a_6+2a_7$\\$-2u_2+u_3)$}&
$\Xi_c^+\to \Xi^0 K^+$&$2a_2+2a_3+a_6-a_7-u_2$\\
\hline

$\Lambda_c^+\to p \pi^0$ & $\frac{\sqrt{2}}{2}(2a_2+2a_3-a_6-a_7+u_2)$ &
$\Xi_c^0\to \Lambda^0 \pi^0$&\makecell[c]{$\frac{\sqrt{3}}{6}(-2a_1-2a_2+4a_3+3a_4-a_5-a_6-a_7$\\$+u_2+u_3)$}\\
\hline

$\Lambda_c^+\to p \eta$ & \makecell[c]{$\frac{\sqrt{2}}{2}c_{\phi}(4a_0+2a_2-2a_3-2a'_0+a_6-a_7+u_2)$\\$+s_{\phi}(-2a_0-2a_1+a'_0+a_4+a_5+a_6-u_3)$} &
$\Xi_c^0\to \Lambda^0 \eta$&\makecell[c]{$\frac{\sqrt{3}}{6}c_{\phi}(12a_0+2a_1+2a_2-4a_3+6a'_0+3a_4+a_5$\\$+a_6+a_7+2u_1+u_2+u_3)$\\$+\frac{\sqrt{6}}{6}s_{\phi}(-6a_0-4a_1-4a_2+2a_3-3a'_0-2a_5$\\$+a_6-2a_7+2u_1)$}\\
\hline

$\Lambda_c^+\to p \eta'$ & \makecell[c]{$\frac{\sqrt{2}}{2}s_{\phi}(4a_0+2a_2-2a_3-2a'_0+a_6-a_7+u_2)$\\$-c_{\phi}(-2a_0-2a_1+a'_0+a_4+a_5+a_6-u_3)$} &
$\Xi_c^0\to \Lambda^0 \eta'$&\makecell[c]{$\frac{\sqrt{3}}{6}s_{\phi}(12a_0+2a_1+2a_2-4a_3+6a'_0+3a_4+a_5$\\$+a_6+a_7+2u_1+u_2+u_3)$\\$-\frac{\sqrt{6}}{6}c_{\phi}(-6a_0-4a_1-4a_2+2a_3-3a'_0-2a_5$\\$+a_6-2a_7+2u_1)$}\\
\hline

$\Lambda_c^+\to n \pi^+$&$2a_2+2a_3+a_6-a_7+u_2$&
$\Xi_c^0\to p K^-$&$-2a_2-a_4-a_7+u_1+u_3$\\
\hline

$\Lambda_c^+\to \Sigma^0 K^+$&$\frac{\sqrt{2}}{2}(2a_1-2a_3-a_4-a_5+u_3)$&
$\Xi_c^0\to n \overline{K}^0$&$2a_1-2a_2-2a_3+a_5-a_7+u_1$\\
\hline

$\Lambda_c^+\to \Sigma^+ K^0$&$2a_1-2a_3+a_4-a_5+u_3$&
$\Xi_c^0\to \Sigma^0 \pi^0$&\makecell[c]{$\frac{1}{2}(2a_1+2a_2-a_4+a_5-a_6+a_7$\\$+2u_1+u_2+u_3)$}
\\
\hline

$\Xi_c^+\to\Lambda^0 \pi^+$& \makecell[c]{$\frac{\sqrt{6}}{6}(2a_1+2a_2-4a_3-3a_4-a_5-a_6-a_7$\\$-u_2-u_3)$}&
$\Xi_c^0\to \Sigma^0 \eta$&\makecell[c]{$\frac{1}{2}c_{\phi}(-4a_0-2a_1-2a_2-2a'_0-a_4-a_5$\\$+a_6-a_7+u_2+u_3)$\\$+\frac{\sqrt{2}}{2}s_{\phi}(2a_0-2a_3+a'_0+a_6)$}\\
\hline

$\Xi_c^+\to p \overline{K}^0$& $2a_1-2a_3+a_4-a_5-u_3$&
$\Xi_c^0\to \Sigma^0 \eta'$&\makecell[c]{$\frac{1}{2}s_{\phi}(-4a_0-2a_1-2a_2-2a'_0-a_4-a_5$\\$+a_6-a_7+u_2+u_3)$\\$-\frac{\sqrt{2}}{2}c_{\phi}(2a_0-2a_3+a'_0+a_6)$}\\
\hline

$\Xi_c^+\to \Sigma^0 \pi^+$&$\frac{\sqrt{2}}{2}(2a_1-2a_2+a_4-a_5+a_6+a_7+u_2-u_3)$&
$\Xi_c^0\to \Xi^0 K^0$&$-2a_1+2a_2+2a_3-a_5+a_7+u_1$\\
\hline

$\Xi_c^+\to \Sigma^+ \pi^0$&$\frac{\sqrt{2}}{2}(-2a_1+2a_2+a_4+a_5+a_6-a_7-u_2+u_3)$&
$\Xi_c^0\to \Sigma^+ \pi^-$&$2a_2+a_4+a_7+u_1+u_3$
\\
\hline

$\Xi_c^+\to \Sigma^+ \eta$&\makecell[c]{$\frac{\sqrt{2}}{2}c_{\phi}(4a_0+2a_1+2a_2-2a'_0-a_4-a_5$\\$-a_6-a_7-u_2-u_3)$\\$+s_{\phi}(-2a_0+2a_3+a'_0-a_6)$}&
$\Xi_c^0\to \Sigma^- \pi^+$&$2a_1+a_5+a_6+u_1+u_2$
\\
\hline

$\Xi_c^+\to \Sigma^+ \eta'$&\makecell[c]{$\frac{\sqrt{2}}{2}s_{\phi}(4a_0+2a_1+2a_2-2a'_0-a_4-a_5$\\$-a_6-a_7-u_2-u_3)$\\$-c_{\phi}(-2a_0+2a_3+a'_0-a_6)$}&
$\Xi_c^0\to \Xi^- K^+$&$-2a_1-a_5-a_6+u_1+u_2$\\
\hline

\end{tabular}
}
\end{center}
\end{table}


Without involving the dynamics of decay details,  decay amplitudes containing SU(3)
breaking effect can be expressed by coefficients of
irreducible representations by expanding Eqs. (\ref{eq:ampSU3}) 
and (\ref{eq:ampSU3b}).  
From the detailed expressions exhibited in 
Tables \ref{tab:amp-cf}, \ref{tab:amp-scs} and \ref{tab:amp-dcs},
the relations of amplitudes (here  $S$-wave amplitudes are taken as an illustration)
between some of the CF and DCS decays can be established, giving
\begin{equation}
\begin{split}
&A(\Lambda_c^+\to \Sigma^0 \pi^+)
=-A(\Lambda_c^+ \to \Sigma^+ \pi^0)\\
&A(\Lambda_c^+\to nK^+)=\sin^2\theta_c A(\Xi_c^+\to \Xi^0 \pi^+)\\
&A(\Xi_c^+\to n\pi^+) = \sin^2\theta_c A(\Lambda_c^+\to \Xi^0 K^+)\\
&A(\Xi_c^+\to \Sigma^+ {K}^0) = \sin^2\theta_c A(\Lambda_c^+\to p \overline{K}^0)\\
&A(\Lambda_c^+\to p {K}^0)= {\sin^2\theta_c} A(\Xi_c^+\to \Sigma^+ \overline{K}^0) \\
&A(\Xi_c^0\to \Sigma^- {K}^+)=-\sin^2\theta_c A(\Xi_c^0\to \Xi^- \pi^+)\\
&A(\Xi_c^0\to p\pi^-)
=-\sin^2\theta_c A(\Xi_c^0\to \Sigma^+ K^-).\label{eq:SU3b}
\end{split}
\end{equation}
Except the first relation which holds between the two CF modes, other connections
are established between CF and DCS modes.
In addition
to the relations in Eq. (\ref{eq:SU3b}),
more relations can be recovered if SU(3) symmetry is strictly respected,\footnote{It can be easily
obtained by setting $u_i$ and $v_i$ zero in Table \ref{tab:amp-scs}.}
\begin{equation}
\begin{split}
&A (\Lambda_c^+\to\Sigma^+ K^0)=A (\Xi_c^+\to p \overline{K}^0)\\
&A(\Lambda_c^+ \to  n \pi^+)=A(\Xi_c^+\to \Xi^0 K^+)\\
&A(\Xi_c^0 \to n \overline{K}^0)=-A(\Xi_c^0 \to \Xi^0 K^0)\\
&A(\Xi_c^0\to p\pi^-)=\sin\theta_c A(\Xi_c^0\to pK^-)
=-\sin\theta_c A(\Xi_c^0\to \Sigma^+ \pi^-)=-\sin^2\theta_c A(\Xi_c^0\to \Sigma^+ K^-)\\
&A(\Xi_c^0\to \Sigma^- {K}^+) =\sin\theta_c A(\Xi_c^0\to \Xi^- K^+)
=-\sin\theta_c A(\Xi_c^0\to \Sigma^- \pi^+)=-\sin^2\theta_c A(\Xi_c^0\to \Xi^- \pi^+)
\end{split}
\label{eq:SU3}
\end{equation}
most of which  manifest in SCS modes. These relations displayed in Eqs. 
(\ref{eq:SU3b}) and (\ref{eq:SU3})
can be confirmed in \cite{Huang:2021aqu} up to a sign in some channels. \footnote{
Note the  sign difference originates from the convention in irreducible representations of 
$\overline{15}$ and $6$, which is unphysical. 
}
It is interesting to notice that any violation of the relations in Eq. (\ref{eq:SU3})
is a signal for  SU(3) symmetry breaking in charmed baryon decays.


\begin{table}[t]
\caption
{The same as Table \ref{tab:amp-cf}, except for doubly Cabibbo-suppressed (DCS) processes.} 
\label{tab:amp-dcs}
\vspace{-0.4cm}
\begin{center}
\renewcommand\arraystretch{1.5}
\resizebox{\textwidth}{!} 
{
\begin{tabular}
{|l c|l c|}
\hline
Channel & $s_c^{-2} A$ & Channel &  $s_c^{-2} A$   \\
\hline
$\Lambda_c^+\to p K^0$&$2a_3-a_4-a_6$&
$\Xi_c^+\to \Sigma^+ K^0$&$-2a_1+a_5+a_6$\\
\hline

$\Lambda_c^+\to n K^+$&$-2a_3-a_4-a_6$&
$\Xi_c^0\to \Lambda^0 K^0$&$\frac{\sqrt{6}}{6}(-2a_1+4a_2+4a_3-a_5-a_6+2a_7)$\\
\hline

$\Xi_c^+\to \Lambda^0 K^+$&$\frac{\sqrt{6}}{6}(-2a_1+4a_2+4a_3+a_5+a_6-2a_7)$&
$\Xi_c^0\to p \pi^-$&$-2a_2-a_4-a_7$\\
\hline

$\Xi_c^+\to p \pi^0$&$\frac{\sqrt{2}}{2}(-2a_2-a_4+a_7)$&
$\Xi_c^0\to n \pi^0$&$\frac{\sqrt{2}}{2}(2a_2-a_4+a_7)$\\
\hline

$\Xi_c^+\to p \eta$&\makecell[c]{$\frac{\sqrt{2}}{2}c_{\phi}(-4a_0-2a_2+2a'_0+a_4+a_7)$\\$+s_{\phi}(2a_0+2a_1-2a_3-a'_0-a_5)$}&
$\Xi_c^0\to n \eta$&\makecell[c]{$\frac{\sqrt{2}}{2}c_{\phi}(-4a_0-2a_2-2a'_0-a_4-a_7)$\\$+s_{\phi}(2a_0+2a_1-2a_3+a'_0+a_5)$}\\
\hline

$\Xi_c^+\to p \eta'$&\makecell[c]{$\frac{\sqrt{2}}{2}s_{\phi}(-4a_0-2a_2+2a'_0+a_4+a_7)$\\$-c_{\phi}(2a_0+2a_1-2a_3-a'_0-a_5)$}&
$\Xi_c^0\to n \eta'$&\makecell[c]{$\frac{\sqrt{2}}{2}s_{\phi}(-4a_0-2a_2-2a'_0-a_4-a_7)$\\$-c_{\phi}(2a_0+2a_1-2a_3+a'_0+a_5)$}\\
\hline

$\Xi_c^+\to n \pi^+$&$-2a_2+a_4+a_7$&
$\Xi_c^0\to \Sigma^0 K^0$&$\frac{\sqrt{2}}{2}(2a_1+a_5-a_6)$\\
\hline

$\Xi_c^+\to \Sigma^0 K^+$&$\frac{\sqrt{2}}{2}(-2a_1+a_5-a_6)$&
$\Xi_c^0\to \Sigma^- K^+$&$-2a_1-a_5-a_6$\\
\hline

\end{tabular}
}
\end{center}
\end{table}


\section{Numerical Analysis}
\label{sec:ana}

\subsection{Fitting schemes}
To fit the existed data and make predictions to the unmeasured channels, 
a $\chi^2$ function is defined by incorporating all the up to date observables, giving 
\begin{equation}
\chi^2=\sum_i \frac{(\mathcal{B}^{\rm{th}}_i-\mathcal{B}_i^{\rm{exp}})^2}{\delta_{1i}^2}
+\sum_i \frac{({\mathcal{R}}^{\rm{th}}_i-{\mathcal{R}}_i^{\rm{exp}})^2}{\delta_{2i}^2}
+\sum_i \frac{({\alpha}^{\rm{th}}_i-{\alpha}_i^{\rm{exp}})^2}{\sigma_i^2},
\end{equation}
in which  $\mathcal{B}^{\rm{th}}, 
\mathcal{R}^{\rm{th}}, \alpha^{\rm{th}}$ ($\mathcal{B}^{\rm{exp}}, 
\mathcal{R}^{\rm{exp}}, \alpha^{\rm{exp}}$) are theoretical formulas 
(experimental central values)
for branching fractions, their ratios and decay asymmetries while
$\delta_{1}, \delta_{2}, \sigma$ are experimental errors for corresponding 
quantities.

\begin{table}[h]
\caption{A summary of current experimental results, including branching fractions, 
branching fraction ratios and decay asymmetries. All the data
without references
is quoted from PDG
\cite{Workman:2022ynf}.
} 
\label{tab:exp}
\vspace{-0.4cm}
\begin{center}
\resizebox{\textwidth}{!} 
{
\begin{tabular}
{| l |c|| l |c|}
\hline
Channel&Expt.&Channel&Expt.\\
\hline
$10^{2}\mathcal{B}(\Lambda_c^+\to\Lambda^0\pi^+)$&$1.30\pm0.07$&
$\mathcal{R}_1=\frac{\mathcal{B}(\Lambda_c^+\to\Sigma^+\eta)}{\mathcal{B}(\Lambda_c^+\to\Sigma^+\pi^0)}$
&$0.25\pm0.03$ \cite{Belle:2022bsi}\\

$10^{2}\mathcal{B}(\Lambda_c^+\to p K_S)$&$1.59\pm0.08$&
$\mathcal{R}_2=\frac{\mathcal{B}(\Lambda_c^+\to\Sigma^+\eta')}{\mathcal{B}(\Lambda_c^+\to\Sigma^+\pi^0)}$
&$0.33\pm0.06$ \cite{Belle:2022bsi}\\

$10^{2}\mathcal{B}(\Lambda_c^+\to \Sigma^0 \pi^+)$&$1.29\pm0.07$&
$\mathcal{R}_3=\frac{\mathcal{B}(\Lambda_c^+\to\Sigma^+\eta')}{\mathcal{B}(\Lambda_c^+\to\Sigma^+\eta)}$&$1.34\pm0.29$ \cite{Belle:2022bsi}\\

$10^{2}\mathcal{B}(\Lambda_c^+\to \Sigma^+ \pi^0)$&$1.25\pm0.10$&
$\mathcal{R}_4=\frac{\mathcal{B}(\Lambda_c^+\to\Lambda^0 K^+)}{\mathcal{B}(\Lambda_c^+\to\Lambda^0\pi^+)}
$&$(4.68\pm0.39)\times 10^{-2}$ \cite{BESIII:2022tnm}\\

$10^{2}\mathcal{B}(\Lambda_c^+\to \Sigma^+ \eta)$&$0.44\pm0.20$&
&$(5.05\pm0.16)\times 10^{-2}$ \cite{Belle:2022uod}\\

&$0.314\pm0.044$ \cite{Belle:2022bsi}&
&$(4.7\pm0.9)\times 10^{-2}$\\

$10^{2}\mathcal{B}(\Lambda_c^+\to \Sigma^+ \eta')$&$1.50\pm0.60$&
$\mathcal{R}_5=\frac{\mathcal{B}(\Lambda_c^+\to\Sigma^0 K^+)}
{\mathcal{B}(\Lambda_c^+\to\Sigma^0\pi^+)}
$&$(3.61\pm0.73)\times 10^{-2}$ \cite{BESIII:2022wxj}\\

&$0.416\pm0.085$ \cite{Belle:2022bsi}&
&$(2.78\pm0.16)\times 10^{-2}$ \cite{Belle:2022uod}\\

$10^{2}\mathcal{B}(\Lambda_c^+\to \Xi^0 K^+)$&$0.55\pm0.07$&
&$(4.0\pm0.6)\times 10^{-2}$\\

$10^{3}\mathcal{B}(\Lambda_c^+\to p \eta)$&$1.42\pm0.12$&
$\mathcal{R}_6=\frac{\mathcal{B}(\Lambda_c^+\to\Sigma^0\pi^+)}{\mathcal{B}(\Lambda_c^+\to\Lambda^0\pi^+)}$&$0.98\pm0.05$\\

$10^{4}\mathcal{B}(\Lambda_c^+\to p \eta')$&$4.73\pm0.97$ \cite{Belle:2021vyq}&
$\mathcal{R}_7=\frac{\mathcal{B}(\Xi_c^0\to\Lambda^0 K_S)}{\mathcal{B}(\Xi_c^0\to\Xi^-\pi^+)}$
&$0.225\pm0.013$\\

&$5.62_{-2.04}^{+2.46}\pm0.26$ \cite{BESIII:2022izy}&
$\mathcal{R}_8=\frac{\mathcal{B}(\Xi_c^0\to\Sigma^0 K_S)}{\mathcal{B}(\Xi_c^0\to\Xi^-\pi^+)}$
&$(3.8\pm0.72)\times 10^{-2}$\\

$10^{4}\mathcal{B}(\Lambda_c^+\to \Lambda^0 K^+)$&$6.21\pm0.61$ \cite{BESIII:2022tnm}&
$\mathcal{R}_9=\frac{\mathcal{B}(\Xi_c^0\to\Sigma^+ K^-)}{\mathcal{B}(\Xi_c^0\to\Xi^-\pi^+)}$
&$0.123\pm0.0122$\\

&$6.57\pm0.40$ \cite{Belle:2022uod}&
$\mathcal{R}_{10}=\frac{\mathcal{B}(\Xi_c^0\to\Xi^-K^+)}{\mathcal{B}(\Xi_c^0\to\Xi^-\pi^+)}$
&$(2.75\pm 0.57)\times 10^{-2}$\\

$10^{4}\mathcal{B}(\Lambda_c^+\to \Sigma^0 K^+)$&$4.7\pm0.95$ \cite{BESIII:2022wxj}&
$\alpha(\Lambda_c^+\to \Lambda^0 \pi^+)$&$-0.84\pm0.09$\\

&$3.58\pm0.28$ \cite{Belle:2022uod}&
&$-0.755\pm0.006$ \cite{Belle:2022uod}\\

$10^{4}\mathcal{B}(\Lambda_c^+\to n \pi^+)$&$6.6\pm1.3$ \cite{BESIII:2022bkj}&
$\alpha(\Lambda_c^+\to p K_S)$&$0.18\pm0.45$\\

$10^{4}\mathcal{B}(\Lambda_c^+\to \Sigma^+ K_S)$&$4.8\pm1.4$ \cite{BESIII:2022wxj}&
$\alpha(\Lambda_c^+\to \Sigma^0 \pi^+)$&$-0.73\pm0.18$\\

$10^{4}\mathcal{B}(\Lambda_c^+\to p\pi^0)$&$<0.80$ \cite{Belle:2021mvw}&
&$-0.463\pm0.018$ \cite{Belle:2022uod}\\

$10^{2}\mathcal{B}(\Xi_c^0\to \Xi^- \pi^+)$&$1.43\pm0.32$&
$\alpha(\Lambda_c^+\to \Sigma^+ \pi^0)$&$-0.55\pm0.11$\\

$10^{3}\mathcal{B}(\Xi_c^0\to \Xi^- K^+)$&$0.38\pm0.12$&
&$-0.48\pm0.03$ \cite{Belle:2022bsi}\\

$10^{3}\mathcal{B}(\Xi_c^0\to \Lambda^0 K_S)$&$3.34\pm0.67$&
$\alpha(\Xi_c^0\to \Xi^- \pi^+)$&$-0.64\pm0.05$\\

$10^{3}\mathcal{B}(\Xi_c^0\to \Sigma^0 K_S)$&$0.69\pm0.24$&
$\alpha(\Lambda_c^+\to \Sigma^+ \eta)$&$-0.99\pm0.06$ \cite{Belle:2022bsi}\\

$10^{3}\mathcal{B}(\Xi_c^0\to\Sigma^+ K^-)$&$1.8\pm0.4$&
$\alpha(\Lambda_c^+\to \Sigma^+ \eta')$&$-0.46\pm0.07$ \cite{Belle:2022bsi}\\

$10^{2}\mathcal{B}(\Xi_c^+\to \Xi^0 \pi^+)$&$1.6\pm0.8$&
$\alpha(\Lambda_c^+\to \Lambda^0 K^+)$&$-0.585\pm0.052$ \cite{Belle:2022uod}\\

&&
$\alpha(\Lambda_c^+\to \Sigma^0 K^+)$&$-0.55\pm0.20$ \cite{Belle:2022uod}\\
\hline
\end{tabular}
}
\end{center}
\end{table}

We use Eq. (\ref{eq:kin}), together with 
explicit amplitudes shown in
Tables \ref{tab:amp-cf}, \ref{tab:amp-scs}, \ref{tab:amp-dcs}
to calculate various theoretical quantities. 
In practice,  two sets of amplitude formulas with or without SU(3) symmetry breaking effects 
are adopted inside Eq.  (\ref{eq:kin}), corresponding to the two different scenarios. 
As for the experimental measurements, in Table \ref{tab:exp} we have collected all the available results up to date, including 20 branching fractions, 10 ratios and 9 decay asymmetries. 
To make a reasonable 
choice of experimental values as part of inputs 
of the $\chi^2$ function,  
we make a comparison among
3 groups of fittings with different choices of experimental data
shown in Appendix  \ref{app:scheme}.

The detailed numerical calculation in Appendix  \ref{app:scheme}
helps  interpret experimental data and make a choice of inputs.
A common feature of all the 6 fitting results is that the one containing SU(3) breaking
has a better goodness than the corresponding fit without symmetry breaking terms.
This SU(3) breaking preference is also indicated from the measurement of $\mathcal{R}_{10}$ 
shown in Table \ref{tab:exp}.  
The branching fraction of 
$\Lambda_c^+\to p \pi^0$ is the main challenge for the first group of fitting (Fit-I and Fit-I'), 
no matter to include SU(3) symmetry breaking or not. To incorporate both $\mathcal{R}_i$
and single decay branching fraction make the observables  double counting, and the values of
$\chi^2_{\rm min}/{\rm d.o.f.}$ are not the best among all the 3 groups as shown in Table \ref{tab:app-br}.
So in Fit-II and Fit-II' all the ratios of branching fractions $\mathcal{R}_i$ are not incorporated
in the definition of $\chi^2$. Moreover, it is easy to find that in
all the 6 fitting schemes the sign of
$\alpha(\Lambda_c\to p K_S)$ is negative while BESIII provided a positive sign before.
The similar situation occurs in the branching fraction of $\Xi_c^+\to \Xi^0\pi^+$.
Considering a correlation between $\Xi_c^+\to \Xi^0\pi^+$ and $\Xi^0\to\Xi^-\pi^+$,
the three observables mentioned above are removed from Fit-II/II', marked as Fit-III/III'.
Apparently, the numbers of  deviations from  experimental results with more than  $1\sigma$ 
is larger  in Fit-II compared with Fit-III.
Though $\mathcal{R}_i$ is not taken as input 
in the definition of $\chi^2$ in the scheme Fit-III, 
more predictions for $\mathcal{R}_i$ are consistent well with experiments.
Therefore, we choose the group of Fit-III and Fit-III' as a
benchmarking combination of input in the following analysis.


\subsection{Results and discussion}
\label{subsec:results}
In total there are  $18$ fitting parameters to be determined in the SU(3) respected situation
while the number increases to $24$ containing SU(3) breaking terms. 
Throughout this work, we rely on the package IMINUIT \cite{iminuit} to perform all the fitting analysis. 
A local minimum of $\chi^2$, from a certain starting point,  can be obtained by combing Newton steps and gradient-descents within the algorithm of the function `Minuit'. To achieve a global minimum, we further take the obtained local
minimum as input 
fixing one of the parameters to different values.
We finally get a set of solution of which most of the parameters can be ensured that its central value locates
at the minimum region.
%

The fitting results of IRA amplitude coefficients are listed in Table \ref{tab:coeff}. In the flavor SU(3) respected framework, 
the coefficients $a_{1,2,3,6}$ and
$b_{2,5}$ are well fitted as their  errors are controlled less than $10\%$ comparing 
with their corresponding central values.
With an around $20\% \sim 50\%$ relative error, $a_{5,7}$ and $b_{1,4,6}$ can provide
useful information. However, the errors are too large 
for $a'_0, a_{0,4}$ and $b'_0, b_{0,3,7}$.
Keeping other $a_i$ similar precision,
the fitting precision of $a_5$ is improved with a price of large uncertainty
for $u_1$ in SU(3) breaking framework.
For the $P$-wave amplitude coefficients, the errors of $b_{1,2,4,6}$ are 
less than $50\%$, while the errors of $b_5, v_{2,3}$ are between $50\%$
and $100\%$. The others are with large uncertainty. 
We should also keep in mind that the errors of each coefficients are also correlated. 
The 
correlation matrix $R$ are given as

\begin{equation}
{\begin{tiny}
R
=\left(\begin{array}{cccccccccccccccccc}
1 & -0.00 & -0.02 & 0.03 & 0.99 & -0.02 & 0.00 & 0.00 & 0.00 & -0.00 & 0.00 & 0.01 & 0.00 & 0.00 & 0.01 & -0.01 & -0.00 & 0.00\\

-0.00 & 1 & 0.09 & 0.09 & 0.00 & -0.01 & 0.89 & -0.13 & 0.11 & 0.00 & -0.25 & 0.02 & 0.09 & -0.00 & -0.09 & -0.67 & -0.05 & 0.07\\

-0.02 & 0.09 & 1 & -0.42 & 0.02 & 0.28 & 0.08 & 0.04 & 0.78 & 0.01 & 0.09 & -0.50 & 0.12 & -0.01 & -0.05 & -0.05 & -0.14 & -0.40\\

0.03 & 0.09 & -0.42 & 1 & -0.03 & -0.39 & -0.05 & -0.39 & -0.10 & -0.00 & 0.22 & 0.17 & 0.24 & 0.00 & -0.01 & 0.09 & -0.26 & 0.27\\

0.99 & 0.00 & 0.02 & -0.03 & 1 & 0.02 & -0.00 & -0.00 & -0.00 & 0.00 & -0.00 & -0.01 & 0.00 & -0.00 & -0.01 & 0.00 & 0.00 & -0.00\\

-0.02 & -0.01 & 0.28 & -0.39 & 0.02 & 1 & -0.11 & -0.37 & -0.19 & 0.01 & 0.17 & -0.20 & 0.09 & -0.01 & 0.12 & 0.03 & -0.13 & -0.25\\

0.00 & 0.89 & 0.08 & -0.05 & -0.00 & -0.11 & 1 & 0.21 & 0.17 & 0.00 & -0.48 & 0.00  & -0.11 & -0.00 & -0.12 & -0.74 & 0.22 & 0.08\\

0.00 & -0.13 & 0.04 & -0.39 & -0.00 & -0.37 & 0.21 & 1 & 0.27 & 0.00 & -0.54 & -0.11 & -0.47 &-0.00 & 0.01 & -0.01 & 0.48 & -0.02\\

0.00 & 0.11 & 0.78 & -0.10 & -0.00 & -0.19 & 0.17 & 0.27 & 1 & -0.00 & -0.02 & -0.41 & 0.10 & 0.00 & -0.13 & -0.13 & -0.04 & -0.29\\

-0.00 & 0.00 & 0.01 & -0.00 & 0.00 & 0.01 & 0.00 & 0.00 & -0.00 & 1 & -0.01 &-0.02 & 0.01 & 0.99 & -0.01 & 0.01 & 0.00 & 0.01\\

0.00 & -0.25 & 0.09 & 0.22 & -0.00 & 0.17 & -0.48 & -0.54 & -0.02 & -0.01 & 1 & 0.00 & 0.78 & 0.01 & 0.03 & 0.38 & -0.89 & -0.07\\

0.00 & 0.02 & -0.50 & 0.17 & -0.01 & -0.20 & 0.00 & -0.11 & -0.41 &-0.02 & 0.00 & 1 & -0.10 & 0.02 & -0.10 & -0.07 & 0.05 & 0.63\\

0.00 & 0.09 & 0.12 & 0.24 & 0.00 & 0.09 & -0.11 & -0.47 & 0.10 & 0.01 & 0.78 & -0.10 & 1 & -0.01 & -0.32 & -0.02 & -0.83 & 0.22\\

0.00 & -0.00 & -0.01 & 0.00 & -0.001 & -0.01 & -0.00 &-0.00 & 0.00 & 0.99 & 0.01 & 0.02 & -0.01 & 1 & 0.01 & -0.01 & -0.00 & -0.01\\

0.01 & -0.09 & -0.05 & -0.01 & -0.01 & 0.12 & -0.12 & 0.01 & -0.13 & -0.01 & 0.03 & -0.10 & -0.32 & 0.01 & 1 & 0.15 & -0.06 & -0.55\\

-0.00 & -0.67 & -0.05 & 0.09 & 0.00 & 0.03 & -0.74 & -0.01 & -0.13 & 0.01 & 0.38 & -0.07 & -0.02 & -0.01 & 0.15 & 1 & -0.14 & -0.01\\

-0.00 & -0.05 & -0.14 & -0.26 & 0.00 & -0.13 & 0.22 & 0.48 & -0.04 & 0.00 & -0.89 & 0.05 & -0.83 & -0.00 & -0.06 & -0.14 & 1 & 0.06\\

0.00 & 0.07 & -0.40 & 0.27 & -0.00 & -0.25  & 0.08 & -0.02 & -0.29 & 0.01 & -0.07 & 0.63 & 0.22 & -0.01 & -0.55 & -0.01 & 0.06 & 1\\
\end{array}\right).
\label{eq:corr}
\end{tiny}
}
\end{equation}
Here we only present the $18\times 18$ correlation matrix in  the exact SU(3) respected scenario as an illustration,  and in practical calculation another $24\times 24$ correlation matrix  for
SU(3) broken case is also needed.

Combing the errors in Table \ref{tab:coeff} and  the correlation matrices, Eq. (\ref{eq:corr}) as an example,  one can straightforwardly calculate
the physical quantities of interest, decay branching fractions and
decay asymmetries of all the two-body weak decays of antitriplet charmed baryons.
The results of the total 16  CF, 26 CSC and 16 DCS channels, together with their $S$- and $P$-wave
amplitudes,  are
displayed in Table \ref{tab:prediction}.
Though 6 more fitting parameters are introduced in the case
of SU(3) breaking, as shown  in Table \ref{tab:app-br} the goodness is better 
for $\chi_{\rm min}^2/{\rm d.o.f.}=1.27$ comparing with
$\chi_{\rm min}^2/{\rm d.o.f.}=2.15$ for the symmetry keeping case.

So far all the CF modes of antitriplet charmed baryons weak decays have been 
measured except the 
channels $\Xi_c^+\to \Sigma^+ K_S$ and $\Xi_c^0\to \Xi^0 \pi^0, \Xi^0\eta, \Xi^0\eta'$.  
The prediction  in current work, 
$\mathcal{B}(\Xi_c^+\to \Sigma^+ K_S)=(0.85^{+0.15}_{-0.16})\times 10^{-2}$ and
corresponding $\alpha=0.65\pm 0.25$, is consistent with the one in \cite{Huang:2021aqu}
but differs the sign of $\alpha$ from the pole model calculation \cite{Zou:2019kzq}.
Suppose that the two modes $\Xi_c^{+,0}\to \Xi^{0,-} \pi^+$ can be
improved by Belle/Belle-II in the near future, we have 
removed them from fitting input lists in current scheme and find other predictions for the CF
measured modes are consistent well  in the two scenarios. To  discriminate 
whether SU(3) symmetry is kept or broken from the aspect of branching fraction,
the CF channel $\Xi_c^0\to \Xi^0 \pi^0$ may play a role as its central value is
around $7$ times larger in the SU(3) breaking case. In addition, the decay asymmetries 
of the yet-to-be measured modes $\Xi_c^0\to \Lambda^0 K_S, \Sigma^0 K_S, \Xi^0\pi^0, \Xi^0\eta, \Xi^0 \eta'$ provide
alternative options to explore the flavor symmetry since the corresponding signs of $\alpha$
differ from each other in the two scenarios. 
Moreover, from the relations exhibited in Eq. (\ref{eq:SU3}) the branching fraction ratios
between corresponding modes should be a good window to observe SU(3) symmetry
in charmed baryon decays. We have defined several ratios as $\mathcal{R}_{1i}$ and 
calculated them based on the fitted parameters in both scenarios presented in 
Table \ref{tab:prediction2}. In principle, once the relative size between $S$- and $P$-wave amplitudes of corresponding channel are known, the ratios can be calculated theoretically.
The predictions of $\mathcal{R}_{1,2, \ldots,10}$ have been mostly confirmed by experimental data from the Fit-III column of Table \ref{tab:app-R}, hence a further 
confirmation of predictions in Table \ref{tab:prediction2} is highly expected.
For the SCS modes, there are only 7 
of 26 
\begin{table}[H]
\caption{ The fitting results of amplitude coefficients in irreducible amplitude approach (see Eq. (\ref{eq:ampSU3}) and  (\ref{eq:ampSU3b}))
in the unit of $10^{-2} G_F {\rm GeV}^2$.} 
\label{tab:coeff}
\vspace{-0.4cm}
\begin{center}
\renewcommand\arraystretch{1.5}
\resizebox{\textwidth}{!} 
{
\begin{tabular}
{| c c c c | c c c c |}
\hline
\multicolumn{4}{|c|}{SU(3) respected (Fit-III')} &
\multicolumn{4}{c|}{SU(3) breaking (Fit-III)}\\
Coefficient & Value & Coefficient & Value &
Coefficient & Value & Coefficient & Value \\
\hline

$a_0$& $-0.10\pm1.00$&
$b_0$& $-0.50\pm2.70$&
$a_0$& $-1.20\pm1.00$&
$b_0$& $-0.70\pm3.30$\\
\hline

$a_1$& $-3.75_{-0.17}^{+0.20}$&
$b_1$& $3.50_{-0.90}^{+1.00}$&
$a_1$& $-3.50_{-0.13}^{+0.21}$&
$b_1$& $8.60_{-0.90}^{+1.30}$\\
\hline

$a_2$& $1.10_{-0.16}^{+0.17}$&
$b_2$& $4.04_{-0.47}^{+0.33}$&
$a_2$& $1.45_{-0.25}^{+0.16}$&
$b_2$& $4.00_{-1.60}^{+0.90}$\\
\hline

$a_3$& $-1.66_{-0.10}^{+0.12}$&
$b_3$& $0.80_{-0.90}^{+1.00}$&
$a_3$& $-1.98_{-0.13}^{+0.15}$&
$b_3$& $-0.80_{-0.70}^{+1.20}$\\
\hline

$a'_0$& $1.60\pm2.00$&
$b'_0$& $2.00\pm5.00$&
$a'_0$& $0.10\pm2.00$&
$b'_0$& $2.00\pm7.00$\\
\hline

$a_4$& $0.04_{-0.21}^{+0.18}$&
$b_4$& $-1.20_{-0.70}^{+0.60}$&
$a_4$& $0.23_{-0.28}^{+0.29}$&
$b_4$& $-3.40\pm0.90$\\
\hline

$a_5$& $1.36_{-0.37}^{+0.45}$&
$b_5$& $-9.10_{-0.70}^{+0.60}$&
$a_5$& $2.14_{-0.23}^{+0.26}$&
$b_5$& $3.00_{-2.20}^{+1.70}$\\
\hline

$a_6$& $1.26_{-0.22}^{+0.16}$&
$b_6$& $8.00_{-1.70}^{+1.60}$&
$a_6$& $1.56_{-0.20}^{+0.14}$&
$b_6$& $10.10_{-2.30}^{+1.50}$\\
\hline

$a_7$& $-1.04_{-0.29}^{+0.33}$&
$b_7$& $0.60_{-1.10}^{+0.90}$&
$a_7$& $-0.59_{-0.42}^{+0.33}$&
$b_7$& $-0.90_{-3.20}^{+1.90}$\\
\hline

&  &
&  &
$u_1$& $0.00\pm110$&
$v_1$& $13.00_{-36}^{+47}$\\
\hline

&  &
&  &
$u_2$& $1.20_{-0.70}^{+0.80}$&
$v_2$& $-3.00_{-2.00}^{+1.70}$\\
\hline

& &
& &
$u_3$& $2.80\pm0.70$&
$v_3$& $4.80_{-1.60}^{+1.40}$\\
\hline
\end{tabular}
}
\end{center}
\end{table}

\noindent branching fractions and 2 of 26 decay asymmetries have been measured,  with them predictions in Fit-III are consistent well. In the case with exact  SU(3) symmetry (Fit-III')  the predicted $\alpha(\Lambda_c^+\to \Sigma^0 K^+)$ has a deviation from Belle measured value \cite{Belle:2022uod} originated mainly from the difference in $P$-wave amplitude.
Among all the 16 DCS channels, the two modes $\Xi_c^+\to \Sigma^0 K^+$ and 
$\Xi_c^+\to \Sigma^+ K^0$  are most possibly to access in next few years
with branching ratio at the order of $10^{-4}$ and negative decay asymmetries in both scenarios,
which are also in agreement  with the theoretical prediction combing the pole model
and MIT bag model \cite{Meng:2020euv}. 

The comparison with experimental values combing fitting results from other groups 
\cite{Geng:2019xbo,Huang:2021aqu,Zhao:2018mov}
as well as the model calculation\cite{Zou:2019kzq, Meng:2020euv},
see Table \ref{tab:comp}, is meaningful.  As a typical example,  
the sign of decay asymmetry 
of $\Lambda_c^+ \to p K_S$
helps  discriminate theoretical calculations, fitting calculations and experimental 
measurement. In all the 6 different fitting schemes of this work 
(see also Table \ref{tab:app-br}),  the predictions provide a negative sign, which
is also supported by the earlier fitting \cite{Geng:2019xbo} as well 
as the model calculation \cite{Zou:2019kzq}. However, the central value of the same
quantity provided by BESIII, together with a previous fitting analysis \cite{Huang:2021aqu}, prefers
a positive sign (though there is still a possibility that the sign is negative considering the errors).
Currently the error is still too large to determine the sign of $\alpha$ yet, when more data accumulated
BESIII, Belle/Belle-II and LHCb are supposed to give the final word. 
The branching fraction of $\Lambda_c^+\to \Sigma^+ \eta'$ was firstly 
measured in 2019 by BESIII measured, given $(1.50\pm 0.60)\times 10^{-2}$ \cite{Workman:2022ynf},  which is supported by earlier  fitting results 
\cite{Geng:2019xbo,Huang:2021aqu}. However,  the recent independent measurement
from Belle gives a smaller

\begin{table}[H]
\caption{The fitting results of all the weak decays of antitriplet charmed baryons  into 
octet pseudoscalar mesons in
both SU(3) flavor symmetry
respected and breaking scenarios, including Cabibbo-favored, singly Cabibbo-suppressed and
doubly Cabibbo-suppressed processes. }
\label{tab:prediction}
\begin{center}
\renewcommand\arraystretch{1}
\resizebox{\textwidth}{!} 
{
\begin{tabular}
{| l | c c c c | c c c c |}
\hline
&\multicolumn{4}{c|}{SU(3) respected}&\multicolumn{4}{c|}{SU(3) broken}\\
\hline
Channel(CF)&
{$A(10^{-1}G_F)$}&{$B(10^{-1}G_F)$}&$10^{2}\mathcal{B}$&{$\alpha$}&
{$A(10^{-1}G_F)$}&{$B(10^{-1}G_F)$}&$10^{2}\mathcal{B}$&{$\alpha$}\\
\hline

$\Lambda_c^+\to \Lambda^0 \pi^+$ & 
$0.26\pm0.01$&$-1.68\pm{0.06}$&
$1.27_{-0.08}^{+0.09}$&$-0.75\pm{0.01}$&
$0.26\pm{0.01}$&$-1.70_{-0.10}^{+0.09}$&
$1.30_{-0.14}^{+0.12}$&$-0.75\pm{0.01}$\\

$\Lambda_c^+\to p \overline{K}^0$ & 
$1.01_{-0.04}^{+0.03}$&$-0.81\pm{0.34}$&
$3.18\pm{0.20}$&$-0.57\pm{0.21}$&
$1.07_{-0.03}^{+0.04}$&$-0.41_{-0.42}^{+0.36}$&
$3.30\pm{0.21}$&$-0.29\pm{0.24}$\\

$\Lambda_c^+\to \Sigma^0 \pi^+$ &
$0.62_{-0.01}^{+0.02}$&$-0.50_{-0.03}^{+0.04}$&
$1.30\pm{0.06}$&$-0.47\pm{0.03}$&
$0.61\pm{0.02}$&$-0.49_{-0.07}^{+0.06}$&
$1.27\pm{0.09}$&$-0.47\pm{0.03}$\\

$\Lambda_c^+\to \Sigma^+ \pi^0$ & 
$-0.62_{-0.01}^{+0.02}$&$0.50_{-0.03}^{+0.04}$&
$1.30\pm{0.06}$&$-0.47\pm{0.03}$&
$-0.61\pm{0.02}$&$0.49_{-0.07}^{+0.06}$&
$1.27\pm{0.09}$&$-0.47\pm{0.03}$\\

$\Lambda_c^+\to \Sigma^+ \eta$  &
$0.21\pm{0.03}$&$-0.99_{-0.12}^{+0.11}$&
$0.314_{-0.051}^{+0.049}$&$-0.96\pm{0.05}$&
$0.19\pm{0.03}$&$-0.98_{-0.15}^{+0.14}$&
$0.297_{-0.067}^{+0.060}$&$-0.95\pm{0.06}$\\

$\Lambda_c^+\to \Sigma^+ \eta'$ &
$0.41\pm{0.04}$&$-0.58\pm{0.11}$&
$0.241_{-0.054}^{+0.052}$&$-0.43\pm{0.07}$&
$0.51\pm{0.05}$&$-0.80\pm{0.15}$&
$0.386_{-0.080}^{+0.079}$&$-0.47\pm{0.05}$\\

$\Lambda_c^+\to \Xi^0 K^+$  & 
$-0.32\pm{0.02}$&$-0.87_{-0.08}^{+0.06}$&
$0.381_{-0.050}^{+0.040}$&$0.91_{-0.04}^{+0.03}$&
$-0.33_{-0.07}^{+0.05}$&$-1.23_{-0.18}^{+0.12}$&
$0.501_{-0.085}^{+0.063}$&$0.99\pm{0.01}$\\

$\Xi_c^+\to \Sigma^+ \overline{K}^0$ & 
$-0.46\pm{0.04}$&$-0.52\pm{0.35}$&
$1.69_{-0.44}^{+0.43}$&$0.65_{-0.31}^{+0.32}$&
$-0.58\pm{0.04}$&$-0.83\pm{0.39}$&
$2.82\pm{0.54}$&$0.77\pm{0.20}$\\

$\Xi_c^+\to \Xi^0 \pi^+$ &
$0.20\pm{0.02}$&$-0.84\pm0.08$&
$0.813_{-0.106}^{+0.111}$&$-0.97\pm{0.03}$&
$0.22\pm{0.04}$&$-0.51_{-0.15}^{+0.14}$&
$0.547_{-0.202}^{+0.195}$&$-0.94\pm{0.05}$\\

$\Xi_c^0\to \Lambda^0 \overline{K}^0$ & 
$0.46_{-0.06}^{+0.07}$&$0.92\pm{0.21}$&
$0.758_{-0.123}^{+0.132}$&$0.94_{-0.10}^{+0.11}$&
$0.39\pm{0.05}$&$-1.01_{-0.29}^{+0.27}$&
$0.668_{-0.196}^{+0.195}$&$-0.996\pm{0.017}$\\

$\Xi_c^0\to \Sigma^0 \overline{K}^0$ &
$0.24\pm{0.04}$&$-0.16_{-0.27}^{+0.26}$&
$0.144_{-0.049}^{+0.051}$&$-0.41_{-0.64}^{+0.63}$&
$0.23_{-0.06}^{+0.05}$&$0.33_{-0.47}^{+0.34}$&
$0.147_{-0.057}^{+0.050}$&$0.76_{-0.39}^{+0.38}$\\

$\Xi_c^0\to \Sigma^+ K^-$ &
$0.12_{-0.06}^{+0.07}$&$0.75_{-0.17}^{+0.13}$&
$0.174_{-0.051}^{+0.041}$&$0.79_{-0.33}^{+0.32}$&
$0.25_{-0.08}^{+0.06}$&$0.37_{-0.62}^{+0.38}$&
$0.186_{-0.071}^{+0.045}$&$0.77_{-0.31}^{+0.27}$\\

$\Xi_c^0\to \Xi^0 \pi^0$ &
$0.20_{-0.05}^{+0.06}$&$0.18\pm{0.11}$&
$0.113_{-0.049}^{+0.059}$&$0.50_{-0.35}^{+0.37}$&
$0.080_{-0.040}^{+0.050}$&$-1.78_{-0.27}^{+0.29}$&
$0.774_{-0.232}^{+0.252}$&$-0.29_{-0.17}^{+0.20}$\\

$\Xi_c^0\to \Xi^0 \eta$ &
$-0.16\pm{0.19}$&$-0.77_{-0.52}^{+0.51}$&
$0.156\pm{0.192}$&$0.97\pm{0.31}$&
$-0.29\pm{0.19}$&$0.58_{-0.70}^{+0.65}$&
$0.243_{-0.290}^{+0.279}$&$-0.84_{-0.51}^{+0.50}$\\

$\Xi_c^0\to \Xi^0 \eta'$ &
$0.20\pm{0.67}$&$0.62\pm{1.77}$&
$0.0683_{-0.3268}^{+0.3272}$&$0.84\pm{2.16}$&
$-0.24_{-0.67}^{+0.66}$&$1.47_{-2.26}^{+2.22}$&
$0.163_{-0.514}^{+0.509}$&$-0.99\pm{0.09}$\\

$\Xi_c^0\to \Xi^- \pi^+$ & 
$-0.49_{-0.07}^{+0.09}$&$0.59_{-0.13}^{+0.14}$&
$0.698_{-0.217}^{+0.248}$&$-0.64\pm{0.07}$&
$-0.33_{-0.05}^{+0.06}$&$3.03_{-0.42}^{+0.38}$&
$2.43_{-0.64}^{+0.60}$&$-0.64_{-0.09}^{+0.11}$\\
\hline

Channel(SCS)&
{$A(10^{-2}G_F)$}&{$B(10^{-2}G_F)$}&$10^{4}\mathcal{B}$&{$\alpha$}&
{$A(10^{-2}G_F)$}&{$B(10^{-2}G_F)$}&$10^{4}\mathcal{B}$&{$\alpha$}\\
\hline
$\Lambda_c^+\to \Lambda^0 K^+$ &
$-1.47\pm{0.04}$&$1.39_{-0.18}^{+0.16}$&
$6.62_{-0.37}^{+0.35}$&$-0.55\pm{0.06}$&
$-1.45\pm{0.05}$&$1.43_{-0.29}^{+0.23}$&
$6.53_{-0.49}^{+0.42}$&$-0.57\pm{0.06}$\\

$\Lambda_c^+\to p \pi^0$ &
$-0.21\pm{0.06}$&$0.17_{-0.52}^{+0.53}$&
$0.157_{-0.088}^{+0.089}$&$-0.60_{-1.57}^{+1.60}$&
$-0.13_{-0.13}^{+0.14}$&$-0.92_{-0.59}^{+0.57}$&
$0.513_{-0.610}^{+0.592}$&$0.61_{-0.25}^{+0.26}$\\

$\Lambda_c^+\to p \eta$ &
$2.22\pm{0.09}$&$0.07\pm{0.91}$&
$13.56_{-1.16}^{+1.10}$&$0.025\pm{0.316}$&
$2.15\pm{0.09}$&$0.06_{-1.11}^{+1.03}$&
$12.75_{-1.06}^{+1.09}$&$0.022\pm{0.328}$\\

$\Lambda_c^+\to p \eta'$ &
$-1.65\pm{0.10}$&$1.89_{-0.35}^{+0.34}$&
$5.93_{-0.73}^{+0.71}$&$-0.63\pm{0.09}$&
$-1.32_{-0.15}^{+0.16}$&$2.63\pm_{-0.52}^{+0.46}$&
$4.65_{-0.77}^{+0.79}$&$-0.89\pm{0.06}$\\

$\Lambda_c^+\to n \pi^+$&
$0.26\pm{0.03}$&$3.83_{-0.18}^{+0.17}$&
$8.15_{-0.69}^{+0.68}$&$0.32\pm{0.05}$&
$0.51_{-0.16}^{+0.19}$&$3.23_{-0.49}^{+0.41}$&
$6.47_{-1.33}^{+1.55}$&$0.67\pm{0.05}$\\

$\Lambda_c^+\to \Sigma^0 K^+$&
$-0.89_{-0.03}^{+0.04}$&$2.49_{-0.13}^{+0.14}$&
$3.56_{-0.22}^{+0.23}$&$-0.97_{-0.01}^{+0.02}$&
$-0.41\pm{0.11}$&$3.81_{-0.26}^{+0.25}$&
$3.71_{-0.36}^{+0.39}$&$-0.67\pm{0.05}$\\

$\Lambda_c^+\to \Sigma^+ K^0$&
$-1.23\pm{0.09}$&$2.99_{-0.33}^{+0.34}$&
$6.22_{-0.67}^{+0.69}$&$-0.93\pm{0.05}$&
$-0.48\pm{0.20}$&$3.86_{-0.48}^{+0.49}$&
$3.98_{-0.92}^{+0.97}$&$-0.74_{-0.10}^{+0.11}$\\

$\Xi_c^+\to\Lambda^0 \pi^+$&
$-0.033_{-0.044}^{+0.045}$&$1.46_{-0.19}^{+0.18}$&
$2.34_{-0.60}^{+0.57}$&$-0.12\pm{0.16}$&
$-0.36\pm{0.21}$&$2.25_{-0.65}^{+0.56}$&
$6.55_{-3.81}^{+3.37}$&$-0.73\pm{0.08}$\\

$\Xi_c^+\to p \overline{K}^0$&
$-1.23\pm{0.09}$&$2.99_{-0.33}^{+0.34}$&
$21.87_{-2.69}^{+2.75}$&$-0.999\pm{0.005}$&
$-1.74\pm{0.15}$&$1.71_{-0.52}^{+0.55}$&
$24.61_{-3.62}^{+3.85}$&$-0.72\pm{0.11}$\\

$\Xi_c^+\to \Sigma^0 \pi^+$&
$-1.71\pm{0.03}$&$2.45_{-0.12}^{+0.11}$&
$28.32_{-1.09}^{+1.05}$&$-0.79\pm{0.02}$&
$-1.98_{-0.17}^{+0.16}$&$0.67_{-0.52}^{+0.39}$&
$30.60_{-5.27}^{+5.17}$&$-0.23\pm{0.04}$\\

$\Xi_c^+\to \Sigma^+ \pi^0$&
$2.13\pm{0.07}$&$-0.29_{-0.57}^{+0.56}$&
$35.12_{-2.16}^{+2.09}$&$-0.094_{-0.187}^{+0.181}$&
$2.54_{-0.15}^{+0.14}$&$1.46_{-0.66}^{+0.57}$&
$52.08_{-6.18}^{+5.86}$&$0.38_{-0.14}^{+0.13}$\\

$\Xi_c^+\to \Sigma^+ \eta$&
$-1.69_{-0.10}^{+0.09}$&$0.84_{-0.88}^{+0.89}$&
$19.90_{-1.92}^{+1.82}$&$-0.31\pm{0.32}$&
$-2.45\pm{0.21}$&$-0.22_{-1.26}^{+1.13}$&
$40.90_{-7.00}^{+6.90}$&$0.058_{-0.227}^{+0.229}$\\

$\Xi_c^+\to \Sigma^+ \eta'$&
$-0.57\pm{0.09}$&$1.67_{-0.34}^{+0.35}$&
$2.28_{-0.73}^{+0.65}$&$-0.94\pm0.07$&
$-1.13\pm{0.21}$&$2.23\pm{0.53}$&
$7.24_{-2.58}^{+2.56}$&$-0.77\pm{0.08}$\\

$\Xi_c^+\to \Xi^0 K^+$&
$0.26\pm{0.03}$&$3.83_{-0.18}^{+0.17}$&
$8.26\pm{0.63}$&$0.47\pm{0.07}$&
$-0.025_{-0.193}^{+0.184}$&$4.58_{-0.59}^{+0.51}$&
$11.11_{-2.89}^{+2.52}$&$-0.039\pm{0.054}$\\

$\Xi_c^0\to \Lambda^0 \pi^0$&
$-0.18_{-0.08}^{+0.09}$&$-0.97_{-0.27}^{+0.28}$&
$0.431_{-0.212}^{+0.205}$&$0.80_{-0.27}^{+0.28}$&
$-0.15_{-0.17}^{+0.16}$&$-3.18_{-0.70}^{+0.60}$&
$3.75_{-1.60}^{+1.36}$&$0.24_{-0.14}^{+0.12}$\\

$\Xi_c^0\to \Lambda^0 \eta$&
$0.79\pm0.51$&$0.31\pm1.40$&
$1.46\pm1.84$&$0.27\pm1.22$&
$0.45\pm{23.69}$&$1.83_{-8.04}^{+10.18}$&
$1.40_{-72.22}^{+74.29}$&$0.94\pm{0.02}$\\

$\Xi_c^0\to \Lambda^0 \eta'$&
$0.21\pm{1.85}$&$1.03_{-4.88}^{+4.87}$&
$0.206\pm{1.432}$&$0.95\pm{5.67}$&
$-0.77\pm6.79$&$4.09_{-6.55}^{+6.69}$&
$3.10_{-15.05}^{+15.54}$&$-0.93\pm{1.87}$\\

$\Xi_c^0\to p K^-$&
$-0.27_{-0.14}^{+0.15}$&$-1.68_{-0.38}^{+0.29}$&
$1.38_{-0.47}^{+0.35}$&$0.65_{-0.32}^{+0.34}$&
$0.058_{-24.571}^{+24.570}$&$3.17_{-8.23}^{+10.45}$&
$4.31_{-23.17}^{+29.53}$&$0.086\pm{0.031}$\\

$\Xi_c^0\to n \overline{K}^0$&
$-0.89_{-0.19}^{+0.22}$&$-2.78_{-0.59}^{+0.54}$&
$5.15_{-1.10}^{+1.18}$&$0.96\pm{0.10}$&
$-0.72_{-24.57}^{+24.56}$&$6.22_{-8.28}^{+10.53}$&
$17.73_{-246.37}^{+257.95}$&$-0.50_{-0.03}^{+0.04}$\\

$\Xi_c^0\to \Sigma^0\pi^0$&
$-0.71_{-0.11}^{+0.12}$&$-0.025_{-0.472}^{+0.447}$&
$1.29_{-0.40}^{+0.45}$&$0.024_{-0.464}^{+0.440}$&
$-0.038_{-24.565}^{+24.566}$&$5.43_{-8.28}^{+10.51}$&
$9.26_{-168.81}^{+176.25}$&$-0.040_{-0.049}^{+0.051}$\\

$\Xi_c^0\to \Sigma^0 \eta$&
$0.90\pm{0.31}$&$0.80\pm{1.01}$&
$1.99_{-1.34}^{+1.35}$&$0.53\pm0.57$&
$1.41_{-0.33}^{+0.32}$&$0.090_{-1.337}^{+1.248}$&
$4.54_{-2.09}^{+2.08}$&$0.041_{-0.456}^{+0.455}$\\

$\Xi_c^0\to \Sigma^0 \eta'$&
$-0.49\pm{1.06}$&$-0.87\pm{2.81}$&
$0.446_{-1.575}^{+1.576}$&$0.71\pm2.12$&
$0.49\pm{1.06}$&$-2.44_{-3.54}^{+3.52}$&
$0.91\pm{2.26}$&$-0.98\pm0.28$\\

$\Xi_c^0\to \Sigma^+ \pi^-$&
$0.27_{-0.14}^{+0.15}$&$1.68_{-0.38}^{+0.29}$&
$1.08_{-0.33}^{+0.25}$&$0.76\pm0.33$&
$1.20\pm{24.57}$&$4.83_{-8.34}^{+10.50}$&
$11.08_{-11.00}^{+16.60}$&$0.95\pm{0.18}$\\

$\Xi_c^0\to \Sigma^- \pi^+$&
$-1.10\pm{0.19}$&$1.32\pm{0.30}$&
$3.66_{-1.15}^{+1.31}$&$-0.71\pm0.07$&
$-0.47\pm24.57$&$9.05_{-8.32}^{+10.55}$&
$25.95_{-241.78}^{+254.15}$&$-0.30_{-0.04}^{+0.05}$\\

$\Xi_c^0\to \Xi^0 K^0$&
$0.89_{-0.19}^{+0.22}$&$2.78_{-0.59}^{+0.54}$&
$3.21_{-0.76}^{+0.67}$&$0.99\pm0.06$&
$0.72_{-24.56}^{+24.57}$&$-0.38_{-8.25}^{+10.43}$&
$1.22_{-38.19}^{+37.82}$&$-0.29_{-0.80}^{+0.77}$\\

$\Xi_c^0\to \Xi^- K^+$&
$1.10_{-0.17}^{+0.19}$&$-1.32\pm{0.30}$&
$3.07_{-0.95}^{+1.09}$&$-0.60\pm0.07$&
$1.01_{-24.58}^{+24.57}$&$-4.56_{-10.38}^{+8.13}$&
$5.99_{-6.68}^{+10.32}$&$-0.98_{-0.26}^{+0.30}$\\
\hline

Channel(DCS)&
{$A(10^{-3}G_F)$}&{$B(10^{-3}G_F)$}&$10^{5}\mathcal{B}$&{$\alpha$}&
{$A(10^{-3}G_F)$}&{$B(10^{-3}G_F)$}&$10^{5}\mathcal{B}$&{$\alpha$}\\
\hline
$\Lambda_c^+\to p K^0$&
$-2.33_{-0.21}^{+0.20}$&$-2.62_{-1.75}^{+1.79}$&
$1.83_{-0.56}^{+0.55}$&$0.74\pm0.32$&
$-2.90_{-0.19}^{+0.20}$&$-4.18_{-1.94}^{+1.99}$&
$3.14_{-0.75}^{+0.76}$&$0.86\pm{0.18}$\\

$\Lambda_c^+\to n K^+$&
$1.02\pm{0.09}$&$-4.23_{-0.39}^{+0.41}$&
$1.08_{-0.15}^{+0.16}$&$-0.89\pm0.05$&
$1.09_{-0.19}^{+0.21}$&$-2.57_{-0.82}^{+0.71}$&
$0.627_{-0.255}^{+0.234}$&$-0.997\pm{0.011}$\\

$\Xi_c^+\to \Lambda^0 K^+$&
$2.05_{-0.06}^{+0.07}$&$2.07_{-0.21}^{+0.23}$&
$3.33_{-0.16}^{+0.19}$&$0.63\pm0.06$&
$2.01_{-0.10}^{+0.08}$&$2.16_{-0.44}^{+0.34}$&
$3.25_{-0.29}^{+0.22}$&$0.67\pm{0.06}$\\

$\Xi_c^+\to p \pi^0$&
$-1.17_{-0.14}^{+0.13}$&$-2.24_{-0.50}^{+0.42}$&
$1.78_{-0.43}^{+0.39}$&$0.99_{-0.04}^{+0.03}$&
$-1.33_{-0.17}^{+0.14}$&$-1.96_{-0.72}^{+0.50}$&
$1.89_{-0.53}^{+0.40}$&$0.92_{-0.10}^{+0.09}$\\

$\Xi_c^+\to p \eta$&
$-2.23\pm{0.12}$&$2.93_{-0.34}^{+0.38}$&
$4.43_{-0.46}^{+0.48}$&$-0.85\pm0.06$&
$-1.97_{-0.13}^{+0.15}$&$2.44_{-0.38}^{+0.48}$&
$3.36_{-0.49}^{+0.53}$&$-0.82_{-0.07}^{+0.08}$\\

$\Xi_c^+\to p \eta'$&
$2.95\pm{0.21}$&$-5.09\pm{0.57}$&
$6.33\pm0.85$&$-0.90\pm0.04$&
$3.39\pm{0.24}$&$-6.08_{-0.82}^{+0.79}$&
$8.52_{-1.28}^{+1.26}$&$-0.92\pm{0.04}$\\

$\Xi_c^+\to n \pi^+$&
$-1.61\pm{0.08}$&$-4.38_{-0.38}^{+0.30}$&
$4.82_{-0.53}^{+0.43}$&$0.98\pm0.02$&
$-1.64_{-0.25}^{+0.20}$&$-6.20_{-0.91}^{+0.62}$&
$7.78_{-1.55}^{+1.03}$&$0.88\pm{0.03}$\\

$\Xi_c^+\to \Sigma^0 K^+$&
$2.71\pm{0.05}$&$-8.59\pm{0.26}$&
$10.56_{-0.43}^{+0.45}$&$-0.9996\pm{0.0007}$&
$2.70_{-0.07}^{+0.08}$&$-8.66_{-0.41}^{+0.42}$&
$10.63_{-0.69}^{+0.70}$&$-0.999\pm{0.001}$\\

$\Xi_c^+\to \Sigma^+ K^0$&
$5.10_{-0.18}^{+0.17}$&$-4.08_{-1.71}^{+1.69}$&
$19.42_{-1.09}^{+1.04}$&$-0.49\pm0.19$&
$5.39_{-0.16}^{+0.18}$&$-2.07_{-2.34}^{+1.60}$&
$20.65_{-1.21}^{+1.30}$&$-0.25\pm{0.20}$\\

$\Xi_c^0\to \Lambda^0 K^0$&
$0.12_{-0.21}^{+0.29}$&$3.02_{-1.10}^{+1.03}$&
$0.273_{-0.194}^{+0.182}$&$0.21_{-0.49}^{+0.53}$&
$0_{-0.32}^{+0.25}$&$-3.97_{-3.10}^{+1.90}$&
$0.468_{-0.555}^{+0.379}$&$0_{-0.37}^{+0.40}$\\

$\Xi_c^0\to p \pi^-$&
$-0.60_{-0.32}^{+0.34}$&$-3.77_{-0.85}^{+0.65}$&
$0.810_{-0.279}^{+0.206}$&$0.63\pm{0.33}$&
$-1.28_{-0.40}^{+0.29}$&$-1.87_{-2.21}^{+1.36}$&
$0.588_{-0.410}^{+0.234}$&$0.92_{-0.24}^{+0.21}$\\

$\Xi_c^0\to n \pi^0$&
$0.40_{-0.22}^{+0.23}$&$3.52_{-0.62}^{+0.78}$&
$0.666_{-0.268}^{+0.215}$&$0.48\pm{0.27}$&
$0.74_{-0.34}^{+0.26}$&$3.74_{-1.64}^{+0.99}$&
$0.845_{-0.764}^{+0.464}$&$0.74_{-0.20}^{+0.19}$\\

$\Xi_c^0\to n \eta$&
$-1.56_{-0.97}^{+0.98}$&$-3.48_{-2.56}^{+2.54}$&
$1.05_{-0.95}^{+0.96}$&$0.999\pm{0.062}$&
$-0.45\pm{0.95}$&$5.80_{-3.40}^{+3.27}$&
$1.43_{-1.65}^{+1.59}$&$-0.36_{-0.71}^{+0.72}$\\

$\Xi_c^0\to n \eta'$&
$-0.35_{-3.38}^{+3.39}$&$-1.09_{-8.94}^{+8.93}$&
$0.0488_{-0.4541}^{+0.4542}$&$0.99\pm4.19$&
$1.71\pm{3.35}$&$-9.85_{-11.29}^{+11.22}$&
$2.77_{-5.50}^{+5.47}$&$-0.78\pm1.02$\\

$\Xi_c^0\to \Sigma^0 K^0$&
$-2.64_{-0.26}^{+0.30}$&$-3.69_{-1.33}^{+1.32}$&
$1.95_{-0.36}^{+0.38}$&$0.74\pm{0.20}$&
$-2.29_{-0.20}^{+0.25}$&$3.60_{-1.51}^{+1.52}$&
$1.55_{-0.35}^{+0.39}$&$-0.81\pm0.17$\\

$\Xi_c^0\to \Sigma^- K^+$&
$2.46_{-0.37}^{+0.43}$&$-2.97_{-0.67}^{+0.68}$&
$1.64_{-0.51}^{+0.58}$&$-0.68\pm{0.07}$&
$1.66_{-0.23}^{+0.31}$&$-15.28_{-2.11}^{+1.93}$&
$6.37_{-1.68}^{+1.58}$&$-0.61_{-0.09}^{+0.11}$\\
\hline
\end{tabular}
}
\end{center}
\end{table}

\begin{table}
\caption{Predictions of branching fraction ratios between some special modes in both
scenarios on SU(3) flavor symmetry. } 
\label{tab:prediction2}
\vspace{-0.2cm}\begin{center}
\renewcommand\arraystretch{1.2}
\begin{tabular}{| l c c |}
\hline
&SU(3) respected&SU(3) broken\\
\hline
$\mR_{11}\equiv \frac{\mathcal{B}(\Lambda_c^+\to \Sigma^+ K^0)}{\mathcal{B}(\Xi_c^+\to p K_S)}$&
$0.57\pm{0.03}$&$0.32\pm{0.09}$\\

$\mR_{12}\equiv \frac{\mathcal{B}(\Lambda_c^+\to n \pi^+)}{\mathcal{B}(\Xi_c^+\to \Xi^0 K^+)}$&
$0.99\pm{0.01}$&$0.58_{-0.24}^{+0.20}$\\

$\mR_{13}\equiv \frac{\mathcal{B}(\Xi_c^0\to n K_S)}{\mathcal{B}(\Xi_c^0\to \Xi^0 K_S)}$&
$1.60_{-0.25}^{+0.26}$&$14.53_{-7931.44}^{+7812.23}$\\

$\mR_{14}\equiv \frac{\mathcal{B}(\Xi_c^0\to p \pi^-)}{\mathcal{B}(\Xi_c^0\to p K^-)}$&
$(5.87\pm{0.07})10^{-2}$&$(1.36_{-18.84}^{+19.62})10^{-2}$\\

$\mR_{15}\equiv \frac{\mathcal{B}(\Xi_c^0\to \Sigma^- K^+)}{\mathcal{B}(\Xi_c^0\to \Xi^- K^+)}$&
$(5.34\pm{0.05})10^{-2}$&$(10.63_{-29.87}^{+46.06})10^{-2}$\\

$\mR_{16}\equiv \frac{\mathcal{B}(\Xi_c^0\to \Xi^- K^+)}{\mathcal{B}(\Xi_c^0\to \Sigma^- \pi^+)}$&
$0.84\pm{0.01}$&$0.23_{-1.15}^{+1.32}$\\
\hline
\end{tabular}
\end{center}
\end{table}

\noindent central value with decreased error, 
$(0.416\pm 0.085)\times 10^{-2}$\cite{Belle:2022bsi}, which locates in the predicted region
in current work.

For the decay of $\Lambda_c^+\to p \eta'$, predictions of the two fittings in SU(3) IRA \cite{Geng:2019xbo,Huang:2021aqu} are both larger than recent measured results
from both BESIII \cite{BESIII:2022izy} and Belle \cite{Belle:2021vyq}, 
while the two scenarios in current work as well as the fitting based on topological
diagram \cite{Zhao:2018mov} can give 
consistent predictions. Although the branching ratio of $\Lambda_c^+\to p \pi^0$
haven't been measured yet, an upper limit has been updated by Belle \cite{Belle:2021mvw},
which is supported also by current fitting in both two scenarios consistent with
the fittings in \cite{Geng:2019xbo} and \cite{Zhao:2018mov}.
For the recent measured decay asymmetry of $\Lambda_c\to \Sigma^0 K^+$ 
\cite{Belle:2022uod},  both current fitting in Fit-III and the pole model calculation \cite{Zou:2019kzq} converge well.

\begin{table}
\caption{The comparison of theoretical predictions for branching fractions
and decay asymmetries in different methods and
the latest measured experimental values. } 
\label{tab:comp}
\vspace{-0.4cm}
\begin{center}
\renewcommand\arraystretch{1.2}
\resizebox{\textwidth}{!} 
{
\begin{tabular}
{|l c c c c c c c|}
\hline
\qquad channel & Fit-III & Fit-III' & GLT \cite{Geng:2019xbo} 
& HXH \cite{Huang:2021aqu} & ZWHY \cite{Zhao:2018mov} 
& ZXMC \cite{Zou:2019kzq, Meng:2020euv}& Expt.  \\
\hline

$10^{2}\mathcal{B}(\Lambda_c^+\to \Lambda^0 \pi^+)$ & 
$1.30_{-0.14}^{+0.12}$&$1.27_{-0.09}^{+0.08}$&
$1.30\pm0.07$&$1.307\pm0.069$& $1.32\pm0.34$&$1.30$ &
$1.30\pm0.07$\\
\hline

$10^{2}\mathcal{B}(\Lambda_c^+\to p K_S)$ & 
$1.65\pm{0.11}$&$1.59\pm{0.10}$&
$1.57\pm0.08$ & $1.587\pm0.077$ & $1.57\pm0.05$&$1.06$&
$1.59\pm0.08$\\
\hline

$10^{2}\mathcal{B}(\Lambda_c^+\to \Sigma^0 \pi^+)$ & 
$1.27\pm{0.09}$&$1.30\pm{0.06}$&
$1.27\pm0.06$ & $1.272\pm0.056$ & $1.26\pm0.32$ &$2.24$&
$1.29\pm0.07$\\
\hline

$10^{2}\mathcal{B}(\Lambda_c^+\to \Sigma^+ \pi^0)$ & 
$1.27\pm{0.09}$&$1.30\pm{0.06}$&
$1.27\pm0.06$ & $1.283\pm0.057$ & $1.23\pm0.17$ &$2.24$&
$1.25\pm0.10$\\
\hline

$10^{2}\mathcal{B}(\Lambda_c^+\to \Sigma^+ \eta)$ &
$0.30_{-0.07}^{+0.06}$&$0.31\pm{0.05}$&
$0.32\pm0.13$ & $0.45\pm0.19$ & $0.47\pm0.22$ &$0.74$&
$0.44\pm0.20$\\
&&&&&&&$0.314\pm0.044$ \cite{Belle:2022bsi}\\
\hline

$10^{2}\mathcal{B}(\Lambda_c^+\to \Sigma^+ \eta')$ & 
$0.39\pm{0.08}$&$0.24\pm{0.05}$&
$1.44\pm0.56$ & $1.5\pm0.60$ & $0.93\pm0.28$ &-&
$1.50\pm0.60$\\
&&&&&&&$0.416\pm0.085$ \cite{Belle:2022bsi}\\
\hline

$10^{2}\mathcal{B}(\Lambda_c^+\to \Xi^0 K^+)$ & 
$0.50_{-0.09}^{+0.06}$&$0.38\pm{0.03}$&
$0.56\pm0.09$ & $0.548\pm0.068$ & $0.59\pm0.17$ &$0.73$&
$0.55\pm0.07$\\
\hline

$10^{3}\mathcal{B}(\Lambda_c^+\to p \eta)$ & 
$1.27\pm{0.11}$&$1.36_{-0.12}^{+0.11}$&
$1.15\pm0.27$ & $1.27\pm0.24$ & $1.14\pm0.35$ &$1.28$&
$1.42\pm0.12$\\
\hline

$10^{4}\mathcal{B}(\Lambda_c^+\to p \eta')$ & 
$4.65_{-0.77}^{+0.79}$&$5.93_{-0.71}^{+0.73}$&
$24.5\pm14.6$ & $27\pm38$ & $7.1\pm1.4$ &-&
$4.73\pm0.97$ \cite{Belle:2021vyq}\\
&&&&&&&$5.62^{+2.46}_{-2.04}\pm0.26$ \cite{BESIII:2022izy}\\
\hline

$10^{4}\mathcal{B}(\Lambda_c^+\to \Lambda^0 K^+)$ & 
$6.54_{-0.49}^{+0.42}$&$6.62_{-0.22}^{+0.23}$&
$6.5\pm1.0$ & $6.4\pm1.0$ & $5.9\pm1.7$ &$10.7$&
$6.21\pm0.61$ \cite{BESIII:2022tnm}\\
&&&&&&&$6.57\pm0.40$ \cite{Belle:2022uod}\\
\hline

$10^{4}\mathcal{B}(\Lambda_c^+\to \Sigma^0 K^+)$ & 
$3.71_{-0.36}^{+0.39}$&$3.56_{-0.69}^{+0.68}$&
$5.4\pm0.7$& $5.04\pm0.56$ & $5.5\pm1.6$ &$7.2$&
$4.7\pm0.95$ \cite{BESIII:2022wxj}\\
&&&&&&&$3.58\pm0.28$ \cite{Belle:2022uod}\\
\hline

$10^{4}\mathcal{B}(\Lambda_c^+\to n \pi^+)$ & 
$6.47_{-1.55}^{+1.33}$&$8.15_{-0.67}^{+0.69}$&
$8.5\pm2.0$ & $3.5\pm1.1$ & $7.7\pm2.0$ &-&
$6.6\pm1.3$ \cite{BESIII:2022bkj}\\
\hline

$10^{4}\mathcal{B}(\Lambda_c^+\to \Sigma^+ K_S)$ & 
$1.99_{-0.46}^{+0.49}$&$3.11_{-0.34}^{+0.35}$&
$5.45\pm0.75$& $1.03\pm0.42$ & $9.55\pm2.4$ &$7.2$&
$4.8\pm1.4$ \cite{BESIII:2022wxj}\\
\hline

$10^{4}\mathcal{B}(\Lambda_c^+\to p\pi^0)$ & 
$0.51_{-0.61}^{+0.59}$&$0.16\pm{0.09}$&
$1.2\pm1.2$& $44.5\pm8.5$ & $0.8_{-0.8}^{+0.9}$ & $1.26$ &
$<0.80$ \cite{Belle:2021mvw}\\
\hline

$10^{2}\mathcal{B}(\Xi_c^0\to \Xi^- \pi^+)$ & 
$2.43_{-0.64}^{+0.60}$&$0.70_{-0.22}^{+0.25}$&
$2.21\pm0.14$& $1.21\pm0.21$ & $1.93\pm0.28$ &$6.47$&
$1.43\pm0.32$\\
\hline

$10^{3}\mathcal{B}(\Xi_c^0\to \Xi^- K^+)$ & 
$0.60_{-0.67}^{+1.03}$&$0.31_{-0.09}^{+0.11}$&
$0.98\pm0.06$& $0.47\pm0.083$ & $0.56\pm0.08$ &$3.90$&
$0.38\pm0.12$\\
\hline

$10^{3}\mathcal{B}(\Xi_c^0\to \Lambda^0 K_S)$ & 
$3.34_{-0.98}^{+0.97}$&$3.79_{-0.61}^{+0.66}$&
$5.25\pm0.3$ & $3.34\pm0.65$ & $4.16\pm2.51$ &$6.65$&
$3.34\pm0.67$\\
\hline

$10^{3}\mathcal{B}(\Xi_c^0\to \Sigma^0 K_S)$ & 
$0.74_{-0.30}^{+0.25}$&$0.73_{-0.25}^{+0.25}$&
$0.4\pm0.4$ & $0.69\pm0.24$ & $3.96\pm0.25$ &$0.2$&
$0.69\pm0.24$\\
\hline

$10^{3}\mathcal{B}(\Xi_c^0\to \Sigma^+ K^-)$ & 
$1.86_{-0.71}^{+0.45}$&$1.74_{-0.51}^{+0.41}$&
$5.9\pm1.1$ & $2.21\pm0.68$ & $22.0\pm5.7$&$4.6$&
$1.8\pm0.4$\\
\hline

$10^{2}\mathcal{B}(\Xi_c^+\to \Xi^0 \pi^+)$ & 
$0.55_{-0.20}^{+0.19}$&$0.81\pm{0.11}$&
$0.38\pm0.20$ & $0.54\pm0.18$ & $0.93\pm0.36$ &$1.72$&
$1.6\pm0.8$\\
\hline

$\alpha(\Lambda_c^+\to \Lambda^0 \pi^+)$ & 
$-0.75\pm{0.01}$&$-0.75\pm{0.01}$&
$-0.87\pm0.10$ & $-0.841\pm0.083$&-&$-0.93$&
$-0.84\pm0.09$\\
&&&&&&&$-0.755\pm0.006$ \cite{Belle:2022uod}\\
\hline

$\alpha(\Lambda_c^+\to p K_S)$ & 
$-0.29\pm{0.24}$&$-0.57\pm{0.21}$&
$-0.90^{+0.22}_{-0.10}$ & $0.19\pm0.41$&-&$-0.75$&
$0.18\pm0.45$\\
\hline

$\alpha(\Lambda_c^+\to \Sigma^0 \pi^+)$ & 
$-0.47\pm{0.03}$&$-0.47\pm{0.03}$&
$-0.35\pm0.27$ & $-0.605\pm0.088$&-&$-0.76$&
$-0.73\pm0.18$\\
&&&&&&&$-0.463\pm0.018$ \cite{Belle:2022uod}\\
\hline

$\alpha(\Lambda_c^+\to \Sigma^+ \pi^0)$ & 
$-0.47\pm{0.03}$&$-0.47\pm{0.03}$&
$-0.35\pm0.27$ & $-0.603\pm0.088$ &-&$-0.76$&
$-0.55\pm0.11$\\
&&&&&&&$-0.48\pm0.03$ \cite{Belle:2022bsi}\\
\hline

$\alpha(\Xi_c^0\to \Xi^- \pi^+)$ & 
$-0.64_{-0.09}^{+0.11}$&$-0.64\pm{0.07}$&
$-0.98^{+0.07}_{-0.02}$& $-0.56\pm0.32$ &-&$-0.95$&
$-0.64\pm0.05$\\
\hline

$\alpha(\Lambda_c^+\to \Sigma^+ \eta)$ & 
$-0.95\pm{0.06}$&$-0.96\pm{0.05}$&
$-0.40\pm0.47$&$0.3\pm3.8$&-&$-0.95$&
$-0.99\pm0.06$ \cite{Belle:2022bsi}\\
\hline

$\alpha(\Lambda_c^+\to \Sigma^+ \eta')$ & 
$-0.47\pm{0.05}$&$-0.43\pm{0.07}$&
$1.00_{-0.17}^{+0.00}$&$0.8\pm1.9$&-&-&
$-0.46\pm0.07$ \cite{Belle:2022bsi}\\
\hline

$\alpha(\Lambda_c^+\to \Lambda^0 K^+)$& 
$-0.57\pm{0.06}$&$-0.55\pm{0.06}$&
$0.32\pm0.32$&$-0.24\pm0.15$&-&$-0.96$&
$-0.585\pm0.052$ \cite{Belle:2022uod}\\
\hline

$\alpha(\Lambda_c^+\to \Sigma^0 K^+)$&
$-0.67\pm{0.05}$&$-0.97_{-0.01}^{+0.02}$&
$-1.00_{-0.00}^{+0.06}$&$-0.953\pm0.040$&-&$-0.73$&
$-0.55\pm0.20$ \cite{Belle:2022uod}\\
\hline

$\alpha(\Lambda_c^+\to \Xi^0 K^+)$&
$0.99\pm{0.01}$&$0.91_{-0.04}^{+0.03}$&
$0.94_{-0.11}^{+0.06}$&$0.866\pm0.090$&-&$0.90$&
\\
\hline
\end{tabular}
}
\end{center}
\end{table}

The dynamical theoretical calculation, the kinetic fitting calculation and the experimental measurement provide
complementary  ways to explore two-body weak decays of charmed baryon
from different aspects.
Each of them can help the other two methodologies to improve their predictions/measured values.
For example, by taking more experimental values into account as inputs, more reliable predictions can
be produced by the fitting methodology.  On the other hand, calculation from both fitting methods and
a model calculation provide a strong motivation to revise $\alpha(\Lambda_c^+\to p K_S)$ in experiment.
Now except $\Xi_c^+\to \Sigma^+ \overline{K}^0$ (or $\Xi_c^+\to \Sigma^+ {K}_S$ ) all the branching fractions of
CF modes as well as part of SCS modes have been measured. It is highly anticipated that
more observables and modes can be accessible, which provide more evidence to confirm
and improve fitting calculation in current work, 
in the coming years.


\section{Summary and conclusion}
\label{sec:con}
In this work, based on SU(3)
flavor symmetry we have carried out a global fitting analysis for the two-body weak 
decays of antitriplet charmed baryons incorporating the latest experimental data.
Two scenarios for fitting, both in the flavor symmetry strictly respected case and the SU(3) breaking case, are considered. In the framework of keeping exact SU(3) symmetry, 
more amplitudes in terms of SU(3) irreducible representation 
have been taken into account benefited from new data.
As for the SU(3) breaking effect, in current working scenario
only SCS processes receive changes while CF and DCS processes 
keep unchanged.

Making use of the data of the measured 20 branching fractions and 9 decay asymmetries,
we obtain the ranges of 18 parameters  in SU(3) keeping scenario as well as the 24
ones in SU(3) breaking case, though some of which are with large uncertainties. 
In general, the goodness is better by incorporating
SU(3) symmetry breaking effect.  In both scenarios, we have calculated 
branching fractions and decay asymmetries for all the CF, SCS and DCS modes
of antitriplet charmed baryon weak decays.  
Most of the predictions for branching fractions and decay asymmetries are 
consistent well with 
experimental measurements. 
For more details, we have 
\begin{itemize}

\item Among the branching fractions of all the 16 CF channels, there are still 4 of them unmeasured.
It is highly anticipated to be measured in the upcoming years for the four modes $\Xi_c^+\to \Sigma^+ K_S, \Xi_c^0\to \Xi^0 \pi^0, \Xi^0 \eta, \Xi^0 \eta'$.

\item Current data prefers 
SU(3) flavor symmetry breaking in charmed baryon weak decays. 

\item  We propose three ways to further explore SU(3) symmetry.
(i) Measuring the branching fraction of CF decay $\Xi_c^0\to \Xi^0\pi^0$ for
the large difference exists in two scenarios. (ii) Measuring
the decay asymmetries in the CF decays $\Xi_c^0\to \Lambda^0 K_S, \Sigma^0 K_S, \Xi^0 \pi^0, \Xi^0 \eta, \Xi^0 \eta'$ as their signs are opposite correspondingly in the two cases. 
(iii) Measuring series of branching fraction ratios, some of which have been presented
in Table \ref{tab:prediction2}.

\item Fittings in different schemes of current work, as well as independent fittings by other groups
all indicate a 2 or 3 times smaller value for the branching fraction of $\Xi_c^+\to \Xi^0 \pi^+$. Hence
we remove this mode as well the correlated mode $\Xi_c^0\to \Xi^- \pi^+$ from the inputs
of fitting. We hope experiments in the near future can help  clarify the two modes.


\item Predictions from most fitting works and pole model calculation prefer a negative
$\alpha(\Lambda_c^+ \to p K_S)$ while BESIII  provided a positive central 
value with large uncertainty. Future measurements from both BESIII and other experiments
are expected to make a further clarification. 

\item Both fittings (in IRA and TDA) and pole model calculation give  a 
positive and large value
for $\alpha(\Lambda_c^+ \to \Xi^0 K^+)$,  then
a check from experiment would be interesting.

\end{itemize}
Together with other results presented in Table \ref{tab:prediction}, all the predictions
of current work
are expected to be checked and improved by the upcoming experiments with the more  accumulated data.


\acknowledgments
The authors benefited by discussions with Prof. H.-Y. Cheng,  C.-Q. Geng and C.-W. Liu.
The communication with Prof. C. D. Lu, X.R. Lyu,  F. -S. Yu,  P.-R. Li
and T. Luo are acknowledged. 
This work is supported by NSFC  under Grant  Nos. U1932104, 12142502 and 12047503,  
and by Guangdong Provincial Key Laboratory of Nuclear Science with No. 2019B121203010.

\vspace{0.5cm}

\noindent \large{\textbf{Note Added}}

\noindent All the authors contribute equally and they are co-first authors, while F. Xu is
the corresponding author.

\newpage

\appendix
\noindent \large{\textbf{Appendix}}
\vspace{-0.2cm}
\section{The comparison among different schemes}
\label{app:scheme}



We present 3 groups of fitting schemes here,
in which those with a prime stand for SU(3) symmetry being strictly respected while
those without a prime receiving symmetry breaking terms. 
In particular, the first group (Fit-I/I') contains all the values of all the observables
in Table \ref{tab:exp} while the second group (Fit-II/II') 
delete all the ratios between branching fractions in the definition of $\chi^2$ in Fit-I/I'.
Then the 

\begin{table}[H]
\caption{Predictions of branching fractions and decay asymmetries in different schemes, 
in which the scheme with a prime stands for the SU(3) respected scenario
while these without a prime containing SU(3) breaking effect. The 
corresponding
experimental values for a comparison are given in the last column.} 
\label{tab:app-br}
\vspace{-0.4cm}
\begin{center}
\renewcommand\arraystretch{1}
\resizebox{\textwidth}{!} 
{
\begin{tabular}
{|l c c c c c c c|}
\hline
\qquad Channel & 
Fit-I & Fit-I' & Fit-II & Fit-II' & Fit-III & Fit-III' & 
Expt.  \\
$\qquad (\chi_{\rm{min}}^2/{\rm{d.o.f.}}$ & 1.16 & 2.60 & 1.98 & 2.20 & 1.27 & 2.15) & \\
\hline 
\hline
$10^{2}\mathcal{B}(\Lambda_c^+\to \Lambda \pi^+)$ & 
$1.32_{-0.14}^{+0.12}$&$1.32_{-0.14}^{+0.12}$&
$1.30_{-0.14}^{+0.12}$&$1.28_{-0.13}^{+0.12}$&
$1.30_{-0.14}^{+0.12}$&$1.27_{-0.09}^{+0.08}$&
$1.30\pm0.07$\\
\hline

$10^{2}\mathcal{B}(\Lambda_c^+\to p K_S)$ & 
$1.59_{-0.38}^{+0.33}$&$1.58_{-0.30}^{+0.25}$&
$1.66_{-0.17}^{+0.15}$&$1.56_{-0.13}^{+0.12}$&
$1.65\pm{0.11}$&$1.59\pm{0.10}$&
$1.59\pm0.08$\\
\hline

$10^{2}\mathcal{B}(\Lambda_c^+\to \Sigma^0 \pi^+)$ & 
$1.28\pm{0.09}$&$1.30\pm{0.09}$&
$1.27\pm{0.09}$&$1.31\pm{0.09}$&
$1.27\pm{0.09}$&$1.30\pm{0.06}$&
$1.29\pm0.07$\\
\hline

$10^{2}\mathcal{B}(\Lambda_c^+\to \Sigma^+ \pi^0)$ & 
$1.28\pm{0.09}$&$1.30\pm{0.09}$&
$1.27\pm{0.09}$&$1.31\pm{0.09}$&
$1.27\pm{0.09}$&$1.30\pm{0.06}$&
$1.25\pm0.10$\\
\hline

$10^{2}\mathcal{B}(\Lambda_c^+\to \Sigma^+ \eta)$ &
$0.30\pm{0.06}$&$0.29\pm{0.06}$&
$0.28\pm{0.06}$&$0.32_{-0.07}^{+0.06}$&
$0.30_{-0.07}^{+0.06}$&$0.31\pm{0.05}$&
$0.44\pm0.20$\\
&&&&&&&
$0.314\pm0.044$ \cite{Belle:2022bsi}\\
\hline

$10^{2}\mathcal{B}(\Lambda_c^+\to \Sigma^+ \eta')$ & 
$0.40\pm{0.08}$&$0.06_{-0.74}^{+0.73}$&
$0.42\pm{0.08}$&$0.26\pm{0.06}$&
$0.39\pm{0.08}$&$0.24\pm{0.05}$&
$1.50\pm0.60$\\
&&&&&&&
$0.416\pm0.085$ \cite{Belle:2022bsi}\\
\hline

$10^{2}\mathcal{B}(\Lambda_c^+\to \Xi^0 K^+)$ & 
$0.60_{-0.09}^{+0.07}$&$0.38_{-0.08}^{+0.06}$&
$0.50_{-0.08}^{+0.06}$&$0.38_{-0.08}^{+0.06}$&
$0.50_{-0.09}^{+0.06}$&$0.38\pm{0.03}$&
$0.55\pm0.07$\\
\hline

$10^{3}\mathcal{B}(\Lambda_c^+\to p \eta)$ & 
$1.31_{-0.25}^{+0.23}$&$1.27_{-0.22}^{+0.21}$&
$1.23_{-0.17}^{+0.16}$&$1.36\pm{0.12}$&
$1.27\pm{0.11}$&$1.36_{-0.12}^{+0.11}$&
$1.42\pm0.12$\\
\hline

$10^{4}\mathcal{B}(\Lambda_c^+\to p \eta')$ & 
$5.04_{-0.78}^{+0.75}$&$3.90_{-0.94}^{+0.97}$&
$4.95\pm{0.77}$&$6.14_{-0.91}^{+0.95}$&
$4.65_{-0.77}^{+0.79}$&$5.93_{-0.71}^{+0.73}$&
$4.73\pm0.97$ \cite{Belle:2021vyq}\\
&&&&&&&
$5.62^{+2.46}_{-2.04}\pm0.26$ \cite{BESIII:2022izy}\\
\hline

$10^{4}\mathcal{B}(\Lambda_c^+\to \Lambda K^+)$ & 
$6.55_{-0.50}^{+0.42}$&$6.62_{-0.49}^{+0.42}$&
$6.68_{-0.50}^{+0.42}$&$6.66_{-0.49}^{+0.42}$&
$6.54_{-0.49}^{+0.42}$&$6.62_{-0.22}^{+0.23}$&
$6.21\pm0.61$ \cite{BESIII:2022tnm}\\
&&&&&&&
$6.57\pm0.40$ \cite{Belle:2022uod}\\
\hline

$10^{4}\mathcal{B}(\Lambda_c^+\to \Sigma^0 K^+)$ & 
$3.63_{-0.36}^{+0.38}$&$3.65_{-0.45}^{+0.49}$&
$3.78_{-0.36}^{+0.39}$&$3.58_{-0.45}^{+0.49}$&
$3.71_{-0.36}^{+0.39}$&$3.56_{-0.69}^{+0.68}$&
$4.7\pm0.95$ \cite{BESIII:2022wxj}\\
&&&&&&&
$3.58\pm0.28$ \cite{Belle:2022uod}\\
\hline

$10^{4}\mathcal{B}(\Lambda_c^+\to n \pi^+)$ & 
$4.67_{-1.38}^{+1.17}$&$8.43_{-1.97}^{+1.66}$&
$5.99_{-1.48}^{+1.28}$&$8.19_{-1.92}^{+1.62}$&
$6.47_{-1.55}^{+1.33}$&$8.15_{-0.67}^{+0.69}$&
$6.6\pm1.3$ \cite{BESIII:2022bkj}\\
\hline

$10^{4}\mathcal{B}(\Lambda_c^+\to \Sigma^+ K_S)$ & 
$1.75_{-0.48}^{+0.50}$&$2.67_{-0.64}^{+0.66}$&
$1.83_{-0.44}^{+0.46}$&$3.14_{-0.72}^{+0.74}$&
$1.99_{-0.46}^{+0.49}$&$3.11_{-0.34}^{+0.35}$&
$4.8\pm1.4$ \cite{BESIII:2022wxj}\\
\hline

$10^{4}\mathcal{B}(\Lambda_c^+\to p\pi^0)$ & 
$2.01_{-1.11}^{+1.08}$&$1.45_{-1.04}^{+1.01}$&
$0.02\pm{0.10}$&$0.15_{-0.17}^{+0.19}$&
$0.51_{-0.61}^{+0.59}$&$0.16\pm{0.09}$&
$<0.80$ \cite{Belle:2021mvw}\\
\hline

$10^{2}\mathcal{B}(\Xi_c^0\to \Xi^- \pi^+)$ & 
$1.40_{-0.28}^{+0.32}$&$1.27_{-0.29}^{+0.32}$&
$1.01_{-0.24}^{+0.27}$&$0.92_{-0.23}^{+0.26}$&
$2.43_{-0.64}^{+0.60}$&$0.70_{-0.22}^{+0.25}$&
$1.43\pm0.32$\\
\hline

$10^{3}\mathcal{B}(\Xi_c^0\to \Xi^- K^+)$ & 
$0.36_{-12.22}^{+12.44}$&$0.56_{-17.01}^{+17.17}$&
$0.39_{-13.95}^{+13.88}$&$0.40_{-14.53}^{+14.65}$&
$0.60_{-0.67}^{+1.03}$&$0.31_{-0.09}^{+0.11}$&
$0.38\pm0.12$\\
\hline

$10^{3}\mathcal{B}(\Xi_c^0\to \Lambda K_S)$ & 
$3.25_{-0.59}^{+0.68}$&$3.15_{-0.58}^{+0.66}$&
$3.96_{-0.68}^{+0.76}$&$4.25_{-0.91}^{+0.93}$&
$3.34_{-0.98}^{+0.97}$&$3.79_{-0.61}^{+0.66}$&
$3.34\pm0.67$\\
\hline

$10^{3}\mathcal{B}(\Xi_c^0\to \Sigma^0 K_S)$ & 
$0.53_{-0.50}^{+0.40}$&$0.57_{-0.80}^{+0.58}$&
$0.78_{-0.40}^{+0.32}$&$0.79_{-0.42}^{+0.31}$&
$0.74_{-0.30}^{+0.25}$&$0.73_{-0.25}^{+0.25}$&
$0.69\pm0.24$\\
\hline

$10^{3}\mathcal{B}(\Xi_c^0\to \Sigma^+ K^-)$ & 
$1.71_{-2.18}^{+1.43}$&$1.51_{-1.08}^{+0.61}$&
$1.74_{-2.03}^{+1.35}$&$1.71_{-2.05}^{+1.22}$&
$1.86_{-0.71}^{+0.45}$&$1.74_{-0.51}^{+0.41}$&
$1.8\pm0.4$\\
\hline

$10^{2}\mathcal{B}(\Xi_c^+\to \Xi^0 \pi^+)$ & 
$0.66\pm{0.21}$&$0.83_{-0.27}^{+0.24}$&
$0.51\pm{0.19}$&$0.83_{-0.26}^{+0.24}$&
$0.55_{-0.20}^{+0.19}$&$0.81\pm{0.11}$&
$1.6\pm0.8$\\
\hline\hline

\hline
$\alpha(\Lambda_c^+\to \Lambda \pi^+)$ & 
$-0.75\pm{0.01}$&$-0.76\pm{0.01}$&
$-0.76\pm{0.01}$&$-0.75\pm{0.01}$&
$-0.75\pm{0.01}$&$-0.75\pm{0.01}$&
$-0.84\pm0.09$\\
&&&&&&&
$-0.755\pm0.006$ \cite{Belle:2022uod}\\
\hline

$\alpha(\Lambda_c^+\to p K_S)$ & 
$-0.95\pm{0.06}$&$-0.98\pm{0.04}$&
$-0.65_{-0.19}^{+0.18}$&$-0.50\pm{0.22}$&
$-0.29\pm{0.24}$&$-0.57\pm{0.21}$&
$0.18\pm0.45$\\
\hline

$\alpha(\Lambda_c^+\to \Sigma^0 \pi^+)$ & 
$-0.47\pm{0.03}$&$-0.47\pm{0.03}$&
$-0.48\pm{0.03}$&$-0.47\pm{0.03}$&
$-0.47\pm{0.03}$&$-0.47\pm{0.03}$&
$-0.73\pm0.18$\\
&&&&&&&
$-0.463\pm0.018$ \cite{Belle:2022uod}\\
\hline

$\alpha(\Lambda_c^+\to \Sigma^+ \pi^0)$ & 
$-0.47\pm{0.03}$&$-0.48\pm{0.03}$&
$-0.48\pm{0.03}$&$-0.47\pm{0.03}$&
$-0.47\pm{0.03}$&$-0.47\pm{0.03}$&
$-0.55\pm0.11$\\
&&&&&&&
$-0.48\pm0.03$ \cite{Belle:2022bsi}\\

\hline

$\alpha(\Xi_c^0\to \Xi^- \pi^+)$ & 
$-0.65\pm{0.05}$&$-0.71\pm{0.05}$&
$-0.65\pm{0.06}$&$-0.66\pm{0.06}$&
$-0.64_{-0.09}^{+0.11}$&$-0.64\pm{0.07}$&
$-0.64\pm0.05$\\
\hline

$\alpha(\Lambda_c^+\to \Sigma^+ \eta)$ & 
$-0.97\pm{0.06}$&$-0.39\pm{0.09}$&
$-0.95\pm{0.06}$&$-0.96\pm{0.05}$&
$-0.95\pm{0.06}$&$-0.96\pm{0.05}$&
$-0.99\pm0.06$ \cite{Belle:2022bsi}\\
\hline

$\alpha(\Lambda_c^+\to \Sigma^+ \eta')$ & 
$-0.50\pm{0.05}$&$-0.88\pm{0.06}$&
$-0.48\pm{0.05}$&$-0.41\pm{0.07}$&
$-0.47\pm{0.05}$&$-0.43\pm{0.07}$&
$-0.46\pm0.07$ \cite{Belle:2022bsi}\\
\hline

$\alpha(\Lambda_c^+\to \Lambda K^+)$& 
$-0.61_{-0.06}^{+0.05}$&$-0.55\pm{0.06}$&
$-0.58_{-0.06}^{+0.05}$&$-0.56\pm{0.06}$&
$-0.57\pm{0.06}$&$-0.55\pm{0.06}$&
$-0.585\pm0.052$ \cite{Belle:2022uod}\\
\hline

$\alpha(\Lambda_c^+\to \Sigma^0 K^+)$&
$-0.63\pm{0.05}$&$-0.98\pm{0.01}$&
$-0.66\pm{0.05}$&$-0.97\pm{0.02}$&
$-0.67\pm{0.05}$&$-0.97_{-0.01}^{+0.02}$&
$-0.55\pm0.20$ \cite{Belle:2022uod}\\
\hline

$\alpha(\Lambda_c^+\to \Xi^0 K^+)$&
$0.98\pm{0.01}$&$0.92_{-0.04}^{+0.03}$&
$0.996_{-0.006}^{+0.005}$&$0.90_{-0.04}^{+0.03}$&
$0.99\pm{0.01}$&$0.91_{-0.04}^{+0.03}$&\\
\hline

\end{tabular}
}
\end{center}
\end{table}

\begin{table}[h!]
\caption{Predictions of ratios between branching fractions in different schemes,
in which the definition of $\mathcal{R}_i$ are given in Table \ref{tab:exp}.}
\label{tab:app-R}
\vspace{-0.4cm}
\begin{center}
\renewcommand\arraystretch{1.2}
\resizebox{\textwidth}{!} 
{
\begin{tabular}
{|l c c c c c c c|}
\hline
Channel & 
Fit-I & Fit-I' & Fit-II & Fit-II' & Fit-III & Fit-III' & 
Expt.  \\
\hline

$\mathcal{R}_1$&
$0.24\pm{0.04}$&$0.22\pm{0.04}$&
$0.22\pm{0.04}$&$0.25_{-0.16}^{+0.06}$&
$0.23_{-0.06}^{+0.05}$&$0.24\pm0.04$&
$0.25\pm0.03$ \cite{Belle:2022bsi}\\
\hline

$\mathcal{R}_2$&
$0.31\pm{0.04}$&$0.05\pm{0.03}$&
$0.33_{-0.06}^{+0.05}$&$0.20_{-0.43}^{+0.04}$&
$0.30_{-0.07}^{+0.06}$&$0.19\pm0.04$&
$0.33\pm0.06$ \cite{Belle:2022bsi}\\
\hline

$\mathcal{R}_3$&
$1.31_{-0.26}^{+0.24}$&$0.20_{-0.24}^{+0.22}$&
$1.49_{-0.36}^{+0.34}$&$0.81_{-1.37}^{+0.14}$&
$1.30_{-0.39}^{+0.38}$&$0.77_{-0.21}^{+0.20}$&
$1.34\pm0.29$ \cite{Belle:2022bsi}\\
\hline

$10^{2}\mathcal{R}_4$&
$4.98_{-0.34}^{+0.30}$&$5.04_{-0.34}^{+0.30}$&
$5.13_{-0.38}^{+0.37}$&$5.22_{-0.40}^{+0.38}$&
$5.03_{-0.65}^{+0.51}$&$5.22_{-0.41}^{+0.39}$&
$4.78\pm0.39$ \cite{BESIII:2022tnm}\\
&&&&&&&
$5.05\pm0.16$ \cite{Belle:2022uod}\\
\hline

$10^{2}\mathcal{R}_5$&
$2.85_{-0.26}^{+0.21}$&$2.81_{-0.17}^{+0.14}$&
$2.98_{-0.18}^{+0.19}$&$2.74\pm{0.13}$&
$2.93_{-0.35}^{+0.31}$&$2.74_{-0.14}^{+0.13}$&
$3.61\pm0.73$ \cite{BESIII:2022wxj}\\
&&&&&&&
$2.78\pm0.16$ \cite{Belle:2022uod}\\
\hline

$\mathcal{R}_6$&
$0.97\pm{0.07}$&$0.99\pm{0.07}$&
$0.97\pm{0.08}$&$1.02\pm{0.08}$&
$0.97_{-0.13}^{+0.11}$&$1.02\pm0.09$&
$0.98\pm0.05$\\
\hline

$\mathcal{R}_7$&
$0.46\pm{0.05}$&$0.49\pm{0.05}$&
$0.79\pm{0.11}$&$0.93_{-0.18}^{+0.19}$&
$0.27_{-0.09}^{+0.05}$&$1.09\pm0.33$&
$0.225\pm0.013$\\
\hline

$10^{2}\mathcal{R}_8$&
$7.55_{-3.58}^{+3.69}$&$8.85_{-2.30}^{+2.90}$&
$15.31_{-7.35}^{+7.54}$&$16.97_{-5.92}^{+6.31}$&
$6.04_{-2.64}^{+2.38}$&$20.58_{-9.13}^{+9.67}$&
$3.8\pm0.7$\\
\hline

$10^{2}\mathcal{R}_9$&
$12.23_{-6.49}^{+5.66}$&$11.89_{-2.34}^{+1.92}$&
$17.28_{-10.08}^{+10.22}$&$18.64_{-6.78}^{+6.13}$&
$7.64_{-3.05}^{+2.49}$&$24.88_{-10.93}^{+10.98}$&
$12.3\pm1.22$\\
\hline

$10^{2}\mathcal{R}_{10}$&
$2.61_{-0.14}^{+0.15}$&$4.38\pm{0.03}$&
$3.90_{-0.18}^{+0.19}$&$4.39\pm{0.02}$&
$2.46_{-2.76}^{+4.26}$&$4.40\pm{0.02}$&
$2.75\pm0.57$\\
\hline
\end{tabular}
}
\end{center}
\end{table}

\noindent third group is defined based on Fit-II/II' by subtracting
$\alpha(\Lambda_c^+\to p K_S)$, $\mathcal{B}(\Xi_c^0\to\Xi^-\pi^+)$
and  $\mathcal{B}(\Xi_c^+\to\Xi^0\pi^+)$.

In Table \ref{tab:app-br} we show predictions of branching fractions and decay
asymmetries in various schemes by  comparing them  with experimental values.
To catch a glimpse of goodness of fit, explicit values of $\chi^2_{\rm{min}}/{\rm{d.o.f.}}$ 
in each scheme are also listed therein.  We also theoretically  calculate the ratios between
particular modes which have been measured as an auxiliary illustration in Table \ref{tab:app-R}.


\bibliographystyle{app/JHEP}
\bibliography{app/reference}

\end{document}